
\magnification=\magstep1


\newbox\SlashedBox
\def\slashed#1{\setbox\SlashedBox=\hbox{#1}
\hbox to 0pt{\hbox to 1\wd\SlashedBox{\hfil/\hfil}\hss}#1}
\def\hboxtosizeof#1#2{\setbox\SlashedBox=\hbox{#1}
\hbox to 1\wd\SlashedBox{#2}}

\def\mathslashed#1{\setbox\SlashedBox=\hbox{$#1$}
\hbox to 0pt{\hbox to 1\wd\SlashedBox{\hfil/\hfil}\hss}#1}

\def\ifsmall{\iffalse}  
\def\titlepagefont{}  

\def\DefineTeXgraphics{%
\special{ps::[global] /TeXgraphics { } def}}  

\def\today{\ifcase\month\or January\or February\or March\or April\or May
\or June\or July\or August\or September\or October\or November\or
December\fi\space\number\day, \number\year}
\def\eatPrefix19{}
\def\Year{\expandafter\eatPrefix\the\year}
\newcount\hours \newcount\minutes
\def\monthname{\ifcase\month\or
January\or February\or March\or April\or May\or June\or July\or
August\or September\or October\or November\or December\fi}
\def\shortmonthname{\ifcase\month\or
Jan\or Feb\or Mar\or Apr\or May\or Jun\or Jul\or
Aug\or Sep\or Oct\or Nov\or Dec\fi}

\def\TimeStamp{\hours\the\time\divide\hours by60%
\minutes -\the\time\divide\minutes by60\multiply\minutes by60%
\advance\minutes by\the\time%
${\rm \shortmonthname}\cdot\if\day<10{}0\fi\the\day\cdot\the\year%
\qquad\the\hours:\if\minutes<10{}0\fi\the\minutes$}







\newif\ifdraftmode
\newif\ifleftlabels  

\def\nolabels{\def\wrlabeL##1{}\def\eqlabeL##1{}\def\reflabeL##1{}}
\def\writelabels{\def\wrlabeL##1{\leavevmode\vadjust{\rlap{\smash%
{\line{{\escapechar=` \hfill\rlap{\sevenrm\hskip.03in\string##1}}}}}}}%
\def\eqlabeL##1{{\escapechar-1\rlap{\sevenrm\hskip.05in\string##1}}}%
\def\reflabeL##1{\noexpand\rlap{\noexpand\sevenrm[\string##1]}}}
\def\writeleftlabels{\def\wrlabeL##1{\leavevmode\vadjust{\rlap{\smash%
{\line{{\escapechar=` \hfill\rlap{\sevenrm\hskip.03in\string##1}}}}}}}%
\def\eqlabeL##1{{\escapechar-1%
\rlap{\sixrm\hskip.05in\string##1}%
\llap{\sevenrm\string##1\hskip.03in\hbox to \hsize{}}}}%
\def\reflabeL##1{\noexpand\rlap{\noexpand\sevenrm[\string##1]}}}
\nolabels

\newdimen\fullhsize
\newdimen\hstitle
\hstitle=\hsize 
\newdimen\hsbody
\hsbody=\hsize 
\newdimen\hbodyoffset
\hbodyoffset=\hoffset 
\newbox\leftpage
\def\abstract#1{#1}
\def\rotated{\special{ps: landscape}
\magnification=1000  
\baselineskip=14pt
\global\hstitle=9truein\global\hsbody=4.75truein
\global\vsize=7truein\global\voffset=-.31truein
\global\hoffset=-0.54in\global\hbodyoffset=-.54truein
\global\fullhsize=10truein
\def\DefineTeXgraphics{%
\special{ps::[global]
/TeXgraphics {currentpoint translate 0.7 0.7 scale
              -80 0.72 mul -1000 0.72 mul translate} def}}
\let\lr=L
\def\ifsmall{\iftrue}
\def\titlepagefont{\twelvepoint}
\trueseventeenpoint
\def\almostshipout##1{\if L\lr \count1=1
      \global\setbox\leftpage=##1 \global\let\lr=R
   \else \count1=2
      \shipout\vbox{\hbox to\fullhsize{\box\leftpage\hfil##1}}
      \global\let\lr=L\fi}

\output={\ifnum\count0=1 
 \shipout\vbox{\hbox to \fullhsize{\hfill\pagebody\hfill}}\advancepageno
 \else
 \almostshipout{\leftline{\vbox{\pagebody\makefootline}}}\advancepageno
 \fi}

\def\abstract##1{{\leftskip=1.5in\rightskip=1.5in ##1\par}} }

\def\linemessage#1{\immediate\write16{#1}}

\global\newcount\secno \global\secno=0
\global\newcount\appno \global\appno=0
\global\newcount\meqno \global\meqno=1
\global\newcount\subsecno \global\subsecno=0
\global\newcount\figno \global\figno=0

\newif\ifAnyCounterChanged
\let\terminator=\relax
\def\normalize#1{\ifx#1\terminator\let\next=\relax\else%
\if#1i\aftergroup i\else\if#1v\aftergroup v\else\if#1x\aftergroup x%
\else\if#1l\aftergroup l\else\if#1c\aftergroup c\else%
\if#1m\aftergroup m\else%
\if#1I\aftergroup I\else\if#1V\aftergroup V\else\if#1X\aftergroup X%
\else\if#1L\aftergroup L\else\if#1C\aftergroup C\else%
\if#1M\aftergroup M\else\aftergroup#1\fi\fi\fi\fi\fi\fi\fi\fi\fi\fi\fi\fi%
\let\next=\normalize\fi%
\next}
\def\makeNormal#1#2{\def\doNormalDef{\edef#1}\begingroup%
\aftergroup\doNormalDef\aftergroup{\normalize#2\terminator\aftergroup}%
\endgroup}

\def\warnIfChanged#1#2{%
\ifundef#1
\else\begingroup%
\edef\oldDefinitionOfCounter{#1}\edef\newDefinitionOfCounter{#2}%
\ifx\oldDefinitionOfCounter\newDefinitionOfCounter%
\else%
\linemessage{Warning: definition of \noexpand#1 has changed.}%
\global\AnyCounterChangedtrue\fi\endgroup\fi}

\def\Section#1{\global\advance\secno by1\relax\global\meqno=1%
\global\subsecno=0%
\bigbreak\bigskip
\centerline{\twelvepoint \bf %
\the\secno. #1}%
\par\nobreak\medskip\nobreak}
\def\tagsection#1{%
\warnIfChanged#1{\the\secno}%
\xdef#1{\the\secno}%
\ifWritingAuxFile\immediate\write\auxfile{\noexpand\xdef\noexpand#1{#1}}\fi%
}
\def\section{\Section}
\def\Subsection#1{\global\advance\subsecno by1\relax\medskip %
\leftline{\bf\the\secno.\the\subsecno\ #1}%
\par\nobreak\smallskip\nobreak}
\def\tagsubsection#1{%
\warnIfChanged#1{\the\secno.\the\subsecno}%
\xdef#1{\the\secno.\the\subsecno}%
\ifWritingAuxFile\immediate\write\auxfile{\noexpand\xdef\noexpand#1{#1}}\fi%
}

\def\subsection{\Subsection}

\def\romappno{\uppercase\expandafter{\romannumeral\appno}}
\def\makeNormalizedRomappno{%
\expandafter\makeNormal\expandafter\normalizedromappno%
\expandafter{\romannumeral\appno}%
\edef\normalizedromappno{\uppercase{\normalizedromappno}}}
\def\Appendix#1{\global\advance\appno by1\relax\global\meqno=1\global\secno=0
\bigbreak\bigskip
\centerline{\twelvepoint \bf Appendix %
\romappno. #1}%
\par\nobreak\medskip\nobreak}
\def\tagappendix#1{\makeNormalizedRomappno%
\warnIfChanged#1{\normalizedromappno}%
\xdef#1{\normalizedromappno}%
\ifWritingAuxFile\immediate\write\auxfile{\noexpand\xdef\noexpand#1{#1}}\fi%
}
\def\appendix{\Appendix}

\def\eqn#1{\makeNormalizedRomappno%
\ifnum\secno>0%
  \warnIfChanged#1{\the\secno.\the\meqno}%
  \eqno(\the\secno.\the\meqno)\xdef#1{\the\secno.\the\meqno}%
     \global\advance\meqno by1
\else\ifnum\appno>0%
  \warnIfChanged#1{\normalizedromappno.\the\meqno}%
  \eqno({\rm\romappno}.\the\meqno)%
      \xdef#1{\normalizedromappno.\the\meqno}%
     \global\advance\meqno by1
\else%
  \warnIfChanged#1{\the\meqno}%
  \eqno(\the\meqno)\xdef#1{\the\meqno}%
     \global\advance\meqno by1
\fi\fi%
\eqlabeL#1%
\ifWritingAuxFile\immediate\write\auxfile{\noexpand\xdef\noexpand#1{#1}}\fi%
}
\def\defeqn#1{\makeNormalizedRomappno%
\ifnum\secno>0%
  \warnIfChanged#1{\the\secno.\the\meqno}%
  \xdef#1{\the\secno.\the\meqno}%
     \global\advance\meqno by1
\else\ifnum\appno>0%
  \warnIfChanged#1{\normalizedromappno.\the\meqno}%
  \xdef#1{\normalizedromappno.\the\meqno}%
     \global\advance\meqno by1
\else%
  \warnIfChanged#1{\the\meqno}%
  \xdef#1{\the\meqno}%
     \global\advance\meqno by1
\fi\fi%
\eqlabeL#1%
\ifWritingAuxFile\immediate\write\auxfile{\noexpand\xdef\noexpand#1{#1}}\fi%
}
\def\anoneqn{\makeNormalizedRomappno%
\ifnum\secno>0
  \eqno(\the\secno.\the\meqno)%
     \global\advance\meqno by1
\else\ifnum\appno>0
  \eqno({\rm\normalizedromappno}.\the\meqno)%
     \global\advance\meqno by1
\else
  \eqno(\the\meqno)%
     \global\advance\meqno by1
\fi\fi%
}
\def\mfig#1#2{\global\advance\figno by1%
\relax#1\the\figno%
\warnIfChanged#2{\the\figno}%
\edef#2{\the\figno}%
\reflabeL#2%
\ifWritingAuxFile\immediate\write\auxfile{\noexpand\xdef\noexpand#2{#2}}\fi%
}

\def\fig#1{\mfig{fig.\ ~}#1}

\catcode`@=11 

\font\ninerm=cmr9
\font\eightrm=cmr8
\font\sixrm=cmr6

\def\loadtrueseventeenpoint{
 \font\seventeenrm=cmr10 at 17.28truept
 \font\seventeeni=cmmi10 at 17.28truept
 \font\seventeenbf=cmbx10 at 17.28truept
 \font\seventeenit=cmti10 at 17.28truept
 \font\seventeensl=cmsl10 at 17.28truept
 \font\seventeensy=cmsy10 at 17.28truept
}
\def\loadfourteenpoint{
\font\fourteenrm=cmr10 at 14.4pt
\font\fourteeni=cmmi10 at 14.4pt
\font\fourteenit=cmti10 at 14.4pt
\font\fourteensl=cmsl10 at 14.4pt
\font\fourteensy=cmsy10 at 14.4pt
\font\fourteenbf=cmbx10 at 14.4pt
}
\def\loadtruetwelvepoint{
\font\twelverm=cmr10 at 12truept
\font\twelvei=cmmi10 at 12truept
\font\twelveit=cmti10 at 12truept
\font\twelvesl=cmsl10 at 12truept
\font\twelvesy=cmsy10 at 12truept
\font\twelvebf=cmbx10 at 12truept
}

\font\ninei=cmmi9
\font\eighti=cmmi8
\font\sixi=cmmi6
\skewchar\ninei='177 \skewchar\eighti='177 \skewchar\sixi='177

\font\ninesy=cmsy9
\font\eightsy=cmsy8
\font\sixsy=cmsy6
\skewchar\ninesy='60 \skewchar\eightsy='60 \skewchar\sixsy='60

\font\ninebf=cmbx9
\font\eightbf=cmbx8
\font\sixbf=cmbx6

\font\ninett=cmtt9
\font\eighttt=cmtt8

\hyphenchar\tentt=-1 
\hyphenchar\ninett=-1
\hyphenchar\eighttt=-1

\font\ninesl=cmsl9
\font\eightsl=cmsl8

\font\nineit=cmti9
\font\eightit=cmti8


\newskip\ttglue
\def\tenpoint{\def\rm{\fam0\tenrm}%
  \textfont0=\tenrm \scriptfont0=\sevenrm \scriptscriptfont0=\fiverm
  \textfont1=\teni \scriptfont1=\seveni \scriptscriptfont1=\fivei
  \textfont2=\tensy \scriptfont2=\sevensy \scriptscriptfont2=\fivesy
  \textfont3=\tenex \scriptfont3=\tenex \scriptscriptfont3=\tenex
  \def\it{\fam\itfam\tenit}\textfont\itfam=\tenit
  \def\sl{\fam\slfam\tensl}\textfont\slfam=\tensl
  \def\bf{\fam\bffam\tenbf}\textfont\bffam=\tenbf \scriptfont\bffam=\sevenbf
  \scriptscriptfont\bffam=\fivebf
  \normalbaselineskip=12pt
  \let\sc=\eightrm
  \let\big=\tenbig
  \setbox\strutbox=\hbox{\vrule height8.5pt depth3.5pt width\z@}%
  \normalbaselines\rm}

\def\twelvepoint{\def\rm{\fam0\twelverm}%
  \textfont0=\twelverm \scriptfont0=\ninerm \scriptscriptfont0=\sevenrm
  \textfont1=\twelvei \scriptfont1=\ninei \scriptscriptfont1=\seveni
  \textfont2=\twelvesy \scriptfont2=\ninesy \scriptscriptfont2=\sevensy
  \textfont3=\tenex \scriptfont3=\tenex \scriptscriptfont3=\tenex
  \def\it{\fam\itfam\twelveit}\textfont\itfam=\twelveit
  \def\sl{\fam\slfam\twelvesl}\textfont\slfam=\twelvesl
  \def\bf{\fam\bffam\twelvebf}\textfont\bffam=\twelvebf
\scriptfont\bffam=\ninebf
  \scriptscriptfont\bffam=\sevenbf
  \normalbaselineskip=12pt
  \let\sc=\eightrm
  \let\big=\tenbig
  \setbox\strutbox=\hbox{\vrule height8.5pt depth3.5pt width\z@}%
  \normalbaselines\rm}

\def\fourteenpoint{\def\rm{\fam0\fourteenrm}%
  \textfont0=\fourteenrm \scriptfont0=\tenrm \scriptscriptfont0=\sevenrm
  \textfont1=\fourteeni \scriptfont1=\teni \scriptscriptfont1=\seveni
  \textfont2=\fourteensy \scriptfont2=\tensy \scriptscriptfont2=\sevensy
  \textfont3=\tenex \scriptfont3=\tenex \scriptscriptfont3=\tenex
  \def\it{\fam\itfam\fourteenit}\textfont\itfam=\fourteenit
  \def\sl{\fam\slfam\fourteensl}\textfont\slfam=\fourteensl
  \def\bf{\fam\bffam\fourteenbf}\textfont\bffam=\fourteenbf%
  \scriptfont\bffam=\tenbf
  \scriptscriptfont\bffam=\sevenbf
  \normalbaselineskip=17pt
  \let\sc=\elevenrm
  \let\big=\tenbig
  \setbox\strutbox=\hbox{\vrule height8.5pt depth3.5pt width\z@}%
  \normalbaselines\rm}

\def\seventeenpoint{\def\rm{\fam0\seventeenrm}%
  \textfont0=\seventeenrm \scriptfont0=\fourteenrm \scriptscriptfont0=\tenrm
  \textfont1=\seventeeni \scriptfont1=\fourteeni \scriptscriptfont1=\teni
  \textfont2=\seventeensy \scriptfont2=\fourteensy \scriptscriptfont2=\tensy
  \textfont3=\tenex \scriptfont3=\tenex \scriptscriptfont3=\tenex
  \def\it{\fam\itfam\seventeenit}\textfont\itfam=\seventeenit
  \def\sl{\fam\slfam\seventeensl}\textfont\slfam=\seventeensl
  \def\bf{\fam\bffam\seventeenbf}\textfont\bffam=\seventeenbf%
  \scriptfont\bffam=\fourteenbf
  \scriptscriptfont\bffam=\twelvebf
  \normalbaselineskip=21pt
  \let\sc=\fourteenrm
  \let\big=\tenbig
  \setbox\strutbox=\hbox{\vrule height 12pt depth 6pt width\z@}%
  \normalbaselines\rm}

\def\ninepoint{\def\rm{\fam0\ninerm}%
  \textfont0=\ninerm \scriptfont0=\sixrm \scriptscriptfont0=\fiverm
  \textfont1=\ninei \scriptfont1=\sixi \scriptscriptfont1=\fivei
  \textfont2=\ninesy \scriptfont2=\sixsy \scriptscriptfont2=\fivesy
  \textfont3=\tenex \scriptfont3=\tenex \scriptscriptfont3=\tenex
  \def\it{\fam\itfam\nineit}\textfont\itfam=\nineit
  \def\sl{\fam\slfam\ninesl}\textfont\slfam=\ninesl
  \def\bf{\fam\bffam\ninebf}\textfont\bffam=\ninebf \scriptfont\bffam=\sixbf
  \scriptscriptfont\bffam=\fivebf
  \normalbaselineskip=11pt
  \let\sc=\sevenrm
  \let\big=\ninebig
  \setbox\strutbox=\hbox{\vrule height8pt depth3pt width\z@}%
  \normalbaselines\rm}

\def\eightpoint{\def\rm{\fam0\eightrm}%
  \textfont0=\eightrm \scriptfont0=\sixrm \scriptscriptfont0=\fiverm%
  \textfont1=\eighti \scriptfont1=\sixi \scriptscriptfont1=\fivei%
  \textfont2=\eightsy \scriptfont2=\sixsy \scriptscriptfont2=\fivesy%
  \textfont3=\tenex \scriptfont3=\tenex \scriptscriptfont3=\tenex%
  \def\it{\fam\itfam\eightit}\textfont\itfam=\eightit%
  \def\sl{\fam\slfam\eightsl}\textfont\slfam=\eightsl%
  \def\bf{\fam\bffam\eightbf}\textfont\bffam=\eightbf \scriptfont\bffam=\sixbf%
  \scriptscriptfont\bffam=\fivebf%
  \normalbaselineskip=9pt%
  \let\sc=\sixrm%
  \let\big=\eightbig%
  \setbox\strutbox=\hbox{\vrule height7pt depth2pt width\z@}%
  \normalbaselines\rm}

\def\tenbig#1{{\hbox{$\left#1\vbox to8.5pt{}\right.\n@space$}}}
\def\ninebig#1{{\hbox{$\textfont0=\tenrm\textfont2=\tensy
  \left#1\vbox to7.25pt{}\right.\n@space$}}}
\def\eightbig#1{{\hbox{$\textfont0=\ninerm\textfont2=\ninesy
  \left#1\vbox to6.5pt{}\right.\n@space$}}}

\def\footnote#1{\edef\@sf{\spacefactor\the\spacefactor}#1\@sf
      \insert\footins\bgroup\eightpoint
      \interlinepenalty100 \let\par=\endgraf
        \leftskip=\z@skip \rightskip=\z@skip
        \splittopskip=10pt plus 1pt minus 1pt \floatingpenalty=20000
        \smallskip\item{#1}\bgroup\strut\aftergroup\@foot\let\next}
\skip\footins=12pt plus 2pt minus 4pt 
\dimen\footins=30pc 

\newinsert\margin
\dimen\margin=\maxdimen

\loadtruetwelvepoint 
\loadtrueseventeenpoint
\catcode`\@=\active
\catcode`@=12  
\catcode`\"=\active

\def\eatOne#1{}
\def\ifundef#1{\expandafter\ifx%
\csname\expandafter\eatOne\string#1\endcsname\relax}
\def\notTrue{\iffalse}\def\isTrue{\iftrue}
\def\ifdef#1{{\ifundef#1%
\aftergroup\notTrue\else\aftergroup\isTrue\fi}}
\def\use#1{\ifundef#1\linemessage{Warning: \string#1 is undefined.}%
{\tt \string#1}\else#1\fi}


\global\newcount\refno \global\refno=1
\newwrite\rfile
\newlinechar=`\^^J
\def\ref#1#2{\the\refno\nref#1{#2}}
\def\nref#1#2{\xdef#1{\the\refno}%
\ifnum\refno=1\immediate\openout\rfile=\jobname.refs\fi%
\immediate\write\rfile{\noexpand\item{[\noexpand#1]\ }#2.}%
\global\advance\refno by1}
\def\lref#1#2{\the\refno\xdef#1{\the\refno}%
\ifnum\refno=1\immediate\openout\rfile=\jobname.refs\fi%
\immediate\write\rfile{\noexpand\item{[\noexpand#1]\ }#2\semi}%
\global\advance\refno by1}
\def\cref#1{\immediate\write\rfile{#1\semi}}

\def\semi{;\hfil\noexpand\break}

\def\listrefs{\vfill\eject\immediate\closeout\rfile
\centerline{{\bf References}}\bigskip\frenchspacing%
\input \jobname.refs\vfill\eject\nonfrenchspacing}

\def\inputAuxIfPresent#1{\immediate\openin1=#1
\ifeof1\message{No file \auxfileName; I'll create one.
}\else\closein1\relax\input\auxfileName\fi%
}
\def\NPB{Nucl.\ Phys.\ B}

\def\PLB{Phys.\ Lett.\ B}

\def\ZPC{Z.\ Phys.\ C}

\newif\ifWritingAuxFile
\newwrite\auxfile
\def\SetUpAuxFile{%
\xdef\auxfileName{\jobname.aux}%
\inputAuxIfPresent{\auxfileName}%
\WritingAuxFiletrue%
\immediate\openout\auxfile=\auxfileName}

\def\L{\left(}\def\R{\right)}
\def\LP{\left.}\def\RP{\right.}
\def\LB{\left[}\def\RB{\right]}

\def\bye{\par\vfill\supereject%
\ifAnyCounterChanged\linemessage{
Some counters have changed.  Re-run tex to fix them up.}\fi%
\end}



\def\Tr{\mathop{\rm Tr}\nolimits}

\def\A#1{{\cal A}_{#1}}

\def\pol{\varepsilon}

\def\ksl{\slashed{k}}

\def\L{\left(}\def\R{\right)}
\def\LP{\left.}\def\RP{\right.}
\def\spa#1.#2{\left\langle#1\,#2\right\rangle}
\def\spb#1.#2{\left[#1\,#2\right]}
\def\lor#1.#2{\left(#1\,#2\right)}
\def\sand#1.#2.#3{%
\left\langle\smash{#1}{\vphantom1}^{-}\right|{#2}%
\left|\smash{#3}{\vphantom1}^{-}\right\rangle}
\def\sandp#1.#2.#3{%
\left\langle\smash{#1}{\vphantom1}^{-}\right|{#2}%
\left|\smash{#3}{\vphantom1}^{+}\right\rangle}
\def\sandpp#1.#2.#3{%
\left\langle\smash{#1}{\vphantom1}^{+}\right|{#2}%
\left|\smash{#3}{\vphantom1}^{+}\right\rangle}
\catcode`@=11  
\def\meqalign#1{\,\vcenter{\openup1\jot\m@th
   \ialign{\strut\hfil$\displaystyle{##}$ && $\displaystyle{{}##}$\hfil
             \crcr#1\crcr}}\,}
\catcode`@=12  

\newread\epsffilein    
\newif\ifepsffileok    
\newif\ifepsfbbfound   
\newif\ifepsfverbose   
\newdimen\epsfxsize    
\newdimen\epsfysize    
\newdimen\epsftsize    
\newdimen\epsfrsize    
\newdimen\epsftmp      
\newdimen\pspoints     
\pspoints=1bp          
\epsfxsize=0pt         
\epsfysize=0pt         
\def\epsfbox#1{\global\def\epsfllx{72}\global\def\epsflly{72}%
   \global\def\epsfurx{540}\global\def\epsfury{720}%
   \def\lbracket{[}\def\testit{#1}\ifx\testit\lbracket
   \let\next=\epsfgetlitbb\else\let\next=\epsfnormal\fi\next{#1}}%
\def\epsfgetlitbb#1#2 #3 #4 #5]#6{\epsfgrab #2 #3 #4 #5 .\\%
   \epsfsetgraph{#6}}%
\def\epsfnormal#1{\epsfgetbb{#1}\epsfsetgraph{#1}}%
\def\epsfgetbb#1{%
%
%
\openin\epsffilein=#1
\ifeof\epsffilein\errmessage{I couldn't open #1, will ignore it}\else
%
%
   {\epsffileoktrue \chardef\other=12
    \def\do##1{\catcode`##1=\other}\dospecials \catcode`\ =10
    \loop
       \read\epsffilein to \epsffileline
       \ifeof\epsffilein\epsffileokfalse\else
%
%
          \expandafter\epsfaux\epsffileline:. \\%
       \fi
   \ifepsffileok\repeat
   \ifepsfbbfound\else
    \ifepsfverbose\message{No bounding box comment in #1; using defaults}\fi\fi
   }\closein\epsffilein\fi}%
%
%
\def\epsfclipstring{}
\def\epsfsetgraph#1{%
   \epsfrsize=\epsfury\pspoints
   \advance\epsfrsize by-\epsflly\pspoints
   \epsftsize=\epsfurx\pspoints
   \advance\epsftsize by-\epsfllx\pspoints
%
%
   \epsfxsize\epsfsize\epsftsize\epsfrsize
   \ifnum\epsfxsize=0 \ifnum\epsfysize=0
      \epsfxsize=\epsftsize \epsfysize=\epsfrsize
      \epsfrsize=0pt
%
%
     \else\epsftmp=\epsftsize \divide\epsftmp\epsfrsize
       \epsfxsize=\epsfysize \multiply\epsfxsize\epsftmp
       \multiply\epsftmp\epsfrsize \advance\epsftsize-\epsftmp
       \epsftmp=\epsfysize
       \loop \advance\epsftsize\epsftsize \divide\epsftmp 2
       \ifnum\epsftmp>0
          \ifnum\epsftsize<\epsfrsize\else
             \advance\epsftsize-\epsfrsize \advance\epsfxsize\epsftmp \fi
       \repeat
       \epsfrsize=0pt
     \fi
   \else \ifnum\epsfysize=0
     \epsftmp=\epsfrsize \divide\epsftmp\epsftsize
     \epsfysize=\epsfxsize \multiply\epsfysize\epsftmp
     \multiply\epsftmp\epsftsize \advance\epsfrsize-\epsftmp
     \epsftmp=\epsfxsize
     \loop \advance\epsfrsize\epsfrsize \divide\epsftmp 2
     \ifnum\epsftmp>0
        \ifnum\epsfrsize<\epsftsize\else
           \advance\epsfrsize-\epsftsize \advance\epsfysize\epsftmp \fi
     \repeat
     \epsfrsize=0pt
    \else
     \epsfrsize=\epsfysize
    \fi
   \fi
%
%
   \ifepsfverbose\message{#1: width=\the\epsfxsize, height=\the\epsfysize}\fi
   \epsftmp=10\epsfxsize \divide\epsftmp\pspoints
   \vbox to\epsfysize{\vfil\hbox to\epsfxsize{%
      \ifnum\epsfrsize=0\relax
        \includegraphics{#1}%
      \else
        \epsfrsize=10\epsfysize \divide\epsfrsize\pspoints
        \includegraphics{#1}%
      \fi
      \hfil}}%
\global\epsfxsize=0pt\global\epsfysize=0pt}%
%
%
{\catcode`\%=12 \global\let\epsfpercent=
%
%
\long\def\epsfaux#1#2:#3\\{\ifx#1\epsfpercent
   \def\testit{#2}\ifx\testit\epsfbblit
      \epsfgrab #3 . . . \\%
      \epsffileokfalse
      \global\epsfbbfoundtrue
   \fi\else\ifx#1\par\else\epsffileokfalse\fi\fi}%
%
%
\def\epsfempty{}%
\def\epsfgrab #1 #2 #3 #4 #5\\{%
\global\def\epsfllx{#1}\ifx\epsfllx\epsfempty
      \epsfgrab #2 #3 #4 #5 .\\\else
   \global\def\epsflly{#2}%
   \global\def\epsfurx{#3}\global\def\epsfury{#4}\fi}%
%
%
\def\epsfsize#1#2{\epsfxsize}
%
%

\epsfverbosetrue

\SetUpAuxFile
\loadfourteenpoint



\def\listrefs{\vskip .3 cm \immediate\closeout\rfile
\noindent{{\bf References}}\smallskip\frenchspacing%
\input \jobname.refs\vfill\eject\nonfrenchspacing}

\def\ref#1#2{\nref#1{#2}}
\overfullrule 0pt
\hfuzz 40pt
\hsize 6. truein
\vsize 8.5 truein

\def\pol{\varepsilon}
\def\eps{\epsilon}

\def\cg{c_\Gamma}

\def\Gr{{\rm Gr}}
\def\Split{\mathop{\rm Split}\nolimits}
\def\Li{\mathop{\rm Li}\nolimits}

\def\tr{{\, \rm tr}}
\def\tn#1#2{t^{[#1]}_{#2}}
\def\treemhv{{\rm tree\ MHV}}

\def\dlips{d{\rm LIPS}}
\def\ksl{\slashed{k}}
\def\qsl{\slashed{q}}

\def\Atree{A^{\rm tree}}

\def\Ram{{\rm R}}
\def\Atree{A^{\rm tree}}
\def\tree{{\rm tree}}

\def\cg{c_\Gamma}

\def\Ls{\mathop{\rm Ls}\nolimits}
\def\Ll{\mathop{\rm L}\nolimits}

\def\e{\epsilon}
\def\hf{{\textstyle{1\over2}}}



\def\ref{\nref}

\ref\ManganoReview{M. Mangano and S.J. Parke, Phys.\ Rep.\ 200:301 (1991)}

\ref\ParkeTaylor{S.J. Parke and T.R. Taylor, Phys.\ Rev.\ Lett.\ 56:2459
(1986)}

\ref\RecursiveA{F.A. Berends and W.T. Giele, Nucl.\ Phys.\ B306:759 (1988)}

\ref\RecursiveB{D.A. Kosower, Nucl.\ Phys.\ B335:23 (1990)}

\ref\SpinorHelicity{
F.A.\ Berends, R.\ Kleiss, P.\ De Causmaecker, R.\ Gastmans and T.\ T.\ Wu,
        Phys.\ Lett.\ 103B:124 (1981)\semi
P.\ De Causmaeker, R.\ Gastmans,  W.\ Troost and  T.T.\ Wu,
Nucl. Phys. B206:53 (1982)\semi
R.\ Kleiss and W.J.\ Stirling,
   Nucl.\ Phys.\ B262:235 (1985)\semi
   J.F.\ Gunion and Z.\ Kunszt, Phys.\ Lett.\ 161B:333 (1985)\semi
 R.\ Gastmans and T.T.\ Wu,
{\it The Ubiquitous Photon: Helicity Method for QED and QCD}
(Clarendon Press,1990)\semi
Z. Xu, D.-H.\ Zhang and L. Chang, Nucl.\ Phys.\ B291:392 (1987)}

\ref\TreeColor{F.A. Berends and W.T. Giele,
Nucl.\ Phys.\ B294:700 (1987)\semi
M.\ Mangano, S. Parke, and Z.\ Xu, Nucl.\ Phys.\ B298:653 (1988)\semi
M.\ Mangano, Nucl.\ Phys.\ B309:461 (1988)}

\ref\ManganoParke{M.\ Mangano and S. J.\ Parke, Nucl.\ Phys.\
B299:673 (1988)}

\ref\Susy{M.T.\ Grisaru, H.N.\ Pendleton and P.\ van Nieuwenhuizen,
Phys. Rev. {D15}:996 (1977)\semi
M.T. Grisaru and H.N. Pendleton, Nucl.\ Phys.\ B124:81 (1977)}

\ref\SusyTree{S.J. Parke and T. Taylor, Phys.\ Lett.\ 157B:81 (1985)\semi
Z. Kunszt, Nucl.\ Phys.\ B271:333 (1986)}

\ref\Ellis{R.K. Ellis and J.C. Sexton, Nucl.\ Phys.\ B269:445 (1986)}

\ref\FiveGluon{Z. Bern, L. Dixon and D.A. Kosower, Phys.\ Rev. Lett.\
70:2677 (1993)}

\ref\StringBased{
Z. Bern and D.A.\ Kosower, Phys.\ Rev.\ Lett.\ 66:1669 (1991);
Nucl.\ Phys.\ B379:451 (1992)\semi
Z. Bern and D.A.\ Kosower, in {\it Proceedings of the PASCOS-91
Symposium}, eds.\ P. Nath and S. Reucroft (World Scientific, 1992)\semi
Z. Bern, Phys.\ Lett.\ 296B:85 (1992)\semi
Z. Bern, D.C. Dunbar and T. Shimada, Phys.\ Lett.\ 312B:277 (1993)}

\ref\Integrals{Z. Bern, L. Dixon and D.A. Kosower,
Phys.\ Lett.\ 302B:299 (1993); erratum B318:649 (1993);
Nucl.\ Phys.\ B412:751 (1994)}

\ref\Mapping{Z. Bern and D.C.\ Dunbar,  Nucl.\ Phys.\ B379:562 (1992)}

\ref\Subsequent{M.J.\ Strassler,  Nucl.\ Phys.\ B385:145 (1992)\semi
C.S.\ Lam, Nucl.\ Phys.\ B397:143 (1993)\semi
M.G.\ Schmidt and C. Schubert, Phys.\ Lett.\ 318B:438 (1993)\semi
D. Fliegner, M.G.\ Schmidt and C. Schubert, HD-THEP-93-44, hep-ph/9401221}

\ref\GN{J.L.\ Gervais and A. Neveu, Nucl.\ Phys.\ B46:381 (1972)}

\ref\Background{G. 't Hooft,
{\it in} Acta Universitatis Wratislavensis no.\
38, 12th Winter School of Theoretical Physics in Karpacz; {\it
Functional and Probabilistic Methods in Quantum Field Theory},
Vol. 1 (1975)\semi
B.S.\ DeWitt, {\it in} Quantum gravity II, eds. C. Isham, R.\ Penrose and
D.\ Sciama (Oxford, 1981)\semi
L.F.\ Abbott, Nucl.\ Phys.\ B185:189 (1981)\semi
L.F. Abbott, M.T. Grisaru and R.K. Schaefer,
Nucl.\ Phys.\ B229:372 (1983)}

\ref\Tasi{
Z. Bern, hep-ph/9304249, in {\it Proceedings of Theoretical
Advanced Study Institute in High Energy Physics (TASI 92)},
eds.\ J. Harvey and J. Polchinski (World Scientific, 1993)}

\ref\WeakInt{
Z.\ Bern and A.\ Morgan, hep-ph/9312218, to appear in Phys.
Rev. D}

\ref\SusyFour{Z. Bern, D. Dunbar, L. Dixon and D. Kosower, preprint
hep-ph/9403226, to appear in Nucl.\ Phys.\ B}

\ref\Color{Z. Bern and D.A.\ Kosower, Nucl.\ Phys.\ B362:389 (1991)}

\ref\KST{Z. Kunszt, A. Signer and Z. Trocsanyi,
Nucl.\ Phys.\ B411:397 (1994)}

\ref\Fermion{Z. Bern, L. Dixon and D.A. Kosower, in preparation}

\ref\BDKconf{Z. Bern, L. Dixon and D.A. Kosower, Proceedings of
Strings 1993,  May 24-29, Berkeley, CA, hep-th/9311026}

\ref\AllPlus{Z. Bern, G. Chalmers, L. Dixon and D.A. Kosower,
Phys.\ Rev.\ Lett.\ 72:2134 (1994)}

\ref\MahlonB{G.D.\ Mahlon, preprint Fermilab-Pub-93/389-T,
hep-ph/9312276}

\ref\TreeCollinear{F.A. Berends and W.T. Giele, Nucl.\ Phys.\ B313:595 (1989)}

\ref\Cutting{L.D.\ Landau, Nucl.\ Phys.\ 13:181 (1959)\semi
 S. Mandelstam, Phys.\ Rev.\ 112:1344 (1958), 115:1741 (1959)\semi
 R.E.\ Cutkosky, J.\ Math.\ Phys.\ 1:429 (1960)}

\ref\MahlonA{G.D.\ Mahlon, Phys.\ Rev.\ D49:2197 (1994)}

\ref\SusyOne{Z. Bern, D. Dunbar, L. Dixon and D. Kosower, in preparation}

\ref\DP{L. Dixon, unpublished}

\ref\ChanPaton{J.E.\ Paton and H.M.\ Chan, Nucl.\ Phys.\ B10:516 (1969)}

\ref\Siegel{W. Siegel, Phys.\ Lett.\ 84B:193 (1979)\semi
D.M.\ Capper, D.R.T.\ Jones and P. van Nieuwenhuizen, Nucl.\ Phys.\
B167:479 (1980)\semi
L.V.\ Avdeev and A.A.\ Vladimirov, Nucl.\ Phys.\ B219:262 (1983)}

\ref\Superspace{S.J. Gates, M.T. Grisaru, M. Rocek and W. Siegel,
 {\it Superspace}, (Benjamin/Cummings, 1983), pages 390-391}

\ref\GSB{M.B.\ Green, J.H.\ Schwarz and L.\ Brink,
 Nucl.\ Phys.\ B198:472 (1982)}

\ref\Morgan{E.W.N.\ Glover and A.G.\ Morgan, Z. Phys.\  C60:175 (1993)}

\ref\Splitting{Z. Bern and G. Chalmers, in preparation}

\ref\PV{L.M.\ Brown and R.P.\ Feynman, Phys.\ Rev.\ 85:231 (1952)\semi
G.\ Passarino and M.\ Veltman, Nucl.\ Phys.\ {B160:151} (1979)\semi
G. 't Hooft and M. Veltman, \NPB{153:365 (1979)}\semi
R.G. Stuart, Comp.\ Phys.\ Comm.\ 48:367 (1988)\semi
R.G. Stuart and A. Gongora, Comp.\ Phys.\ Comm.\ 56:337 (1990)}

\ref\MVNV{
D.B. Melrose, Il Nuovo Cimento 40A:181 (1965)\semi
W. van Neerven and J.A.M. Vermaseren, Phys.\ Lett.\ 137B:241 (1984)\semi
G.J. van Oldenborgh and J.A.M. Vermaseren, \ZPC{46:425 (1990)}\semi
G.J. van Oldenborgh, PhD thesis, University of Amsterdam (1990)\semi
A. Aeppli, PhD thesis, University of Zurich (1992)\semi
A. Denner, U. Nierste, and R. Scharf,
  \NPB{367:637 (1991)}\semi
N.I. Usyukina and A.I. Davydychev, \PLB{298:363 (1993)}; \PLB{305:136 (1993)}}

\ref\Lewin{L.\ Lewin, {\it Dilogarithms and Associated Functions\/}
(Macdonald, 1958)}

\ref\MahlonP{G. Mahlon, private communication}

\ref\Roland{K. Roland, Phys.\ Lett.\ 289B:148 (1992)\semi
G. Cristofano, R. Marotta and K. Roland, Nucl.\ Phys.\ B392:345 (1993)\semi
M.G. Schmidt and C. Schubert, hep-th/9403158}


\noindent
hep-ph/9405248\hfill SLAC--PUB--6490

\hfill Saclay/SPhT--T94/055

\hfill UCLA/94/TEP/17

\hskip .3 cm


\baselineskip 12 pt
\centerline{\bf ONE-LOOP GAUGE THEORY AMPLITUDES WITH}
\centerline{{\bf AN ARBITRARY NUMBER OF EXTERNAL LEGS}%
\footnote{${}^*$}%
{Talk presented by Z.B. at Continuous Advances in QCD,
Minneapolis, Feb.\ 18-20, 1994 } }

\vskip .3 cm
\centerline{\ninerm ZVI BERN}
\baselineskip=13pt
\centerline{\nineit Department of Physics, UCLA, Los Angeles, CA 90024}
\vglue 0.3cm

\centerline{\ninerm LANCE DIXON}
\centerline{\nineit Stanford Linear Accelerator Center, Stanford University,
Stanford, CA 94309}
\vglue 0.3cm

\centerline{\ninerm DAVID C. DUNBAR}
\baselineskip=13pt
\centerline{\nineit Department of Physics, UCLA, Los Angeles, CA 90024}

\vglue 0.2cm
\centerline{\ninerm and}
\vglue 0.2cm
\centerline{\ninerm DAVID A. KOSOWER}
\baselineskip12truept
\centerline{\nineit Service de Physique Th\'eorique de Saclay,
 Centre d'Etudes de Saclay}
\centerline{\nineit F-91191 Gif-sur-Yvette cedex, France}

\vglue 0.7cm
\centerline{\tenrm ABSTRACT}
\vglue 0.3cm
{\rightskip=3pc
\leftskip=3pc
\tenrm\baselineskip=12pt
\noindent
We review recent progress in calculations of one-loop QCD amplitudes.
By imposing the consistency requirements of unitarity and correct behavior
as the momenta of two legs become collinear, we construct ans\"atze for
one-loop amplitudes with an arbitrary number of external legs.  For
supersymmetric amplitudes, which can be thought of as components of
QCD amplitudes, the cuts uniquely specify the amplitude.}

\baselineskip 16. pt

\vglue 0.5cm

\noindent
{\bf 1. Introduction.}

\vskip .1 cm

Multi-jet processes at colliders require knowledge of matrix elements
with multiple final state partons.  The discovery of new physics
relies to a large extent on the subtraction of known QCD physics from
the data.  Unfortunately, perturbative QCD amplitudes are notoriously
difficult to calculate even at tree level~[\use\ManganoReview].  It
has nevertheless been possible to derive a set of extremely simple
formulae at tree level for ``maximally helicity-violating'' (MHV)
amplitudes with an arbitrary number of external gluons
[\use\ParkeTaylor,\use\RecursiveA,\use\RecursiveB].

Three main ideas contributed to progress at
tree-level: the use of a spinor helicity basis~[\use\SpinorHelicity] for gluon
polarization vectors; the color decomposition of the amplitudes
[\use\TreeColor,\use\ManganoParke];
and a recursive technique for calculating
the kinematical coefficients of the different color factors
[\use\RecursiveA,\use\RecursiveB].
Supersymmetry Ward identities [\use\Susy] have also
proven useful [\use\SusyTree].

In 1986, R. K.\ Ellis and J. Sexton defined the state of the art for
one-loop computations by their calculation of the four-parton matrix
elements [\use\Ellis]. The next step beyond this, the calculation of
five-point amplitudes, is significantly more complicated and seemed
unobtainable with standard techniques.  As an example of the
complexity, consider the pentagon diagram one would encounter in a
brute-force five-gluon computation.  A naive count of the number of
terms gives approximately $6^5$ terms since each non-abelian vertex
contains six terms.  (This count is reduced by the use of on-shell
conditions but increased since each internal momentum is a sum of
external momenta and the loop momentum.)  Each term is associated
with an integral which evaluates to an expression on the order of a
page in length.  This means that one is faced with the order of $10^4$
pages of algebra for this diagram alone.

Last year, all one-loop five-gluon helicity amplitudes
[\use\FiveGluon] were obtained by use of string-based techniques
[\use\StringBased] coupled with improvements in integration
[\use\Integrals] and spinor helicity methods.  This method helps
minimize the algebra by reducing the size of the initial expressions,
as well as organizing the results into smaller gauge invariant pieces,
called partial amplitudes.  This leads to much smaller intermediate
expressions.  Much of the simplification can be understood in
terms of field theory [\use\Mapping,\use\Subsequent]. Some of the
ingredients which improve the computational efficiency beyond the
tree-level techniques, are better gauge choices
[\use\GN,\use\Background,\use\Mapping], a supersymmetric decomposition
of amplitudes [\use\FiveGluon,\use\Tasi,\use\WeakInt,\use\SusyFour]
and a relation which allows one to obtain all subleading-color partial
amplitudes [\use\Color] in terms of leading-color amplitudes
[\use\SusyFour].  These three ideas were motivated by string theory
but may also be derived in field theory.
Supersymmetry Ward identities provide an additional helpful tool at
one loop [\use\KST].  These types of simplifications can also be used
for external fermions [\use\Fermion]. An explicit example of one such
amplitude will be provided here.

Here we will discuss exact one-loop results in QCD for particular
helicity configurations and an arbitrary number of external legs
[\use\BDKconf,\use\AllPlus,\use\MahlonB,\use\SusyFour].
The three additional ingredients which enter are:

\item{1)} Consistency of amplitudes as two legs become collinear
[\use\ParkeTaylor,\use\TreeCollinear]. At tree level, the collinear
(and soft) behavior of amplitudes has been widely used as a
consistency check; the same technique can be applied at one-loop
[\use\AllPlus,\use\SusyFour].
Using the collinear limits as a consistency
condition, one can construct guesses for higher-point amplitudes
based on previously calculated amplitudes.

\item{2)} Unitarity, in the form of the Cutkosky rules [\use\Cutting]
which fix the cuts of any loop amplitude in terms of tree
amplitudes.  It is often easier to reconstruct the loop amplitudes from
the cuts than to calculate the full amplitude directly; unitarity
therefore provides a powerful
tool for obtaining parts of amplitudes containing cuts.

\item{3)} A one-loop unitarity result which states that an amplitude
is uniquely determined by its cuts if the $m$-point loop integrals
contributing to it have at most $m-2$ powers of the loop momentum
in the numerator; i.e., if the diagrams lead to integrands with
two powers less than the maximum possible in gauge theory.
Examples of such amplitudes are provided by supersymmetric
gauge theories.

\noindent
Mahlon has also introduced recursive techniques for constructing
one-loop amplitudes [\use\MahlonA,\use\MahlonB], which are complementary
to these.

Why do we need compact analytic results when the matrix elements are
ultimately inserted into numerical programs?  Without compact results,
numerical instabilities generically arise from the vanishing of
spurious Gram determinant denominators in the expression; such
spuriously singular denominators have been removed from the results
which we present.  Compact analytic results also make it easier to
compare independent calculations.  Finally, compact analytic results
are crucial for extensions to an arbitrary number of external legs.

For the identical-helicity case, we used the collinear constraints
to construct an ansatz~[\use\BDKconf,\use\AllPlus] for an
arbitrary number of external legs which was later proven correct by a
recursive procedure~[\use\MahlonB].  Mahlon has also constructed an
all-$n$ formula for the configuration with one leg of opposite
helicity from the rest [\use\MahlonB]. Since tree-level amplitudes
vanish for these helicity configurations, these loop amplitudes
do not contribute to multi-jet QCD cross-sections at next-to-leading
order in $\alpha_s$.
Amplitudes with two opposite-helicity gluons are
more useful in this regard.  For these helicity configurations,
we have obtained results
for an arbitrary number of external legs for both $N=4$
[\use\SusyFour] and $N=1$ supersymmetric amplitudes [\use\SusyOne].
 When using a supersymmetric decomposition of
amplitudes discussed in
refs.~[\use\FiveGluon,\use\Tasi,\use\WeakInt,\use\SusyFour] (as described in
the next section), these may
be thought of as two of the three terms in an $n$-gluon
QCD amplitude.  The third component is the contribution of a scalar in
the loop, for which
unitarity will not give the complete result and which is more difficult to
obtain.

Other related examples of loop amplitudes that are known
for all $n$ include the $n$-photon massless QED amplitudes where all
photon helicities are identical, or all but one are identical, which
have recently been shown to vanish for five or more legs by
Mahlon~[\use\MahlonA].  The QED results can be generalized to
amplitudes with external photons and gluons, interacting via a
massless quark loop.  Amplitudes with five or more legs, where three
or more legs are photons instead of gluons, also vanish
when all the helicities are identical~[\use\AllPlus], and when
one of the photon helicities is reversed~[\use\DP].

\vskip .2 cm
\noindent
{\bf 2. Review of Previous Results.}

\vskip .1 cm

Tree-level amplitudes for $U(N_c)$ or $SU(N_c)$ gauge theory
with $n$ external gluons can be decomposed
into color-ordered partial amplitudes, multiplied by
an associated color trace [\use\ChanPaton,\use\TreeColor,\ManganoParke].
Summing over all non-cyclic permutations reconstructs the full
amplitude $\A{n}^\tree$ from the partial amplitudes $A_n^\tree(\sigma)$,
$$
\A{n}^\tree(\{k_i,\lambda_i,a_i\}) =
g^{n-2} \sum_{\sigma\in S_n/Z_n} \Tr(T^{a_{\sigma(1)}}
\cdots T^{a_{\sigma(n)}})
\ A_n^\tree(k_{\sigma(1)}^{\lambda_{\sigma(1)}},\ldots,
            k_{\sigma(n)}^{\lambda_{\sigma(n)}})\ ,
\eqn\TreeAmplitudeDecomposition
$$
where $k_i$, $\lambda_i$, and $a_i$ are respectively the momentum,
helicity ($\pm$), and color index of the $i$-th external
gluon, $g$ is the coupling constant, and $S_n/Z_n$ is the set of
non-cyclic permutations of $\{1,\ldots, n\}$.
The $U(N_c)$ ($SU(N_c)$) generators $T^a$ are the set of hermitian
(traceless hermitian) $N_c\times N_c$ matrices,
normalized so that $\Tr\L T^a T^b\R = \delta^{ab}$.
The color decomposition~(\use\TreeAmplitudeDecomposition)
can be derived in conventional field theory simply by using
$
f^{abc} = -i \Tr\L \LB T^a, T^b\RB T^c\R/\sqrt2,
$
where the $T^a$ may by either $SU(N_c)$ matrices or $U(N_c)$ matrices.

In a supersymmetric theory, amplitudes with all helicities identical,
or all but one identical, vanish due to supersymmetry Ward
identities~[\use\Susy].
Tree-level gluon amplitudes in super-Yang-Mills and in
purely gluonic Yang-Mills are identical
(fermions do not appear at this order), so that
$$
A_n^\tree(1^{\pm},2^{+}, \ldots,n^+) = 0.
\eqn\TreeVanish
$$
Parity may of course be used to simultaneously reverse all helicities
in a partial amplitude.
The non-vanishing Parke-Taylor formul\ae~[\use\ParkeTaylor] are for
maximally helicity-violating (MHV) partial amplitudes,
those with two negative helicities and the rest positive,
$$
\eqalign{
  A_{jk}^\treemhv(1,2,\ldots,n)\
&=\ i\, { {\spa{j}.{k}}^4 \over \spa1.2\spa2.3\cdots\spa{n}.1 }\ ,
  \cr}
\eqn\PT
$$
where we have introduced the notation
$$
\eqalign{
  A_{jk}^{\rm MHV}(1,2,\ldots,n)\ &\equiv\
  A_n(1^+,\ldots,j^-,\ldots,k^-,
                \ldots,n^+),  \cr}
\eqn\mhvdef
$$
for a partial amplitude where $j$ and $k$ are the only legs with
negative helicity. Our convention is that all legs are outgoing.
The result~(\PT) is written in terms of spinor inner-products,
$\spa{j}.{l} = \langle j^- | l^+ \rangle = \bar{u}_-(k_j) u_+(k_l)$ and
$\spb{j}.{l} = \langle j^+ | l^- \rangle = \bar{u}_+(k_j) u_-(k_l)$,
where $u_\pm(k)$ is a massless Weyl spinor with momentum $k$ and
chirality $\pm$~[\use\SpinorHelicity,\use\ManganoReview].

For one-loop amplitudes, one may perform a similar color decomposition
to the tree-level decomposition (\use\TreeAmplitudeDecomposition);
in this case, there are up to two traces
over color matrices [\use\Color],
and one must also sum over the different spins $J$ of the internal
particles circulating in the loop.
When all internal particles transform as color adjoints,
the result takes the form
$$
{\cal A}_n\L \{k_i,\lambda_i,a_i\}\R =
g^n  \sum_{J} n_J\,\sum_{c=1}^{\lfloor{n/2}\rfloor+1}
      \sum_{\sigma \in S_n/S_{n;c}}
     \Gr_{n;c}\L \sigma \R\,A_{n;c}^{[J]}(\sigma),
\eqn\ColorDecomposition
$$
where ${\lfloor{x}\rfloor}$ is the largest integer less than or equal to $x$
and $n_J$ is the number of particles of spin $J$.
The leading color-structure factor,
$$
\Gr_{n;1}(1) = N_c\ \Tr\L T^{a_1}\cdots T^{a_n}\R \, ,
\anoneqn
$$
is just $N_c$ times the tree color factor, and the subleading color
structures ($c>1)$ are given by
$$
\Gr_{n;c}(1) = \Tr\L T^{a_1}\cdots T^{a_{c-1}}\R\,
\Tr\L T^{a_c}\cdots T^{a_n}\R \, .
\anoneqn
$$
$S_n$ is the set of all permutations of $n$ objects,
and $S_{n;c}$ is the subset leaving $\Gr_{n;c}$ invariant.
Once again it is convenient to use $U(N_c)$ matrices; the extra $U(1)$
decouples from all final results [\use\Color].
(For internal particles in the fundamental ($N_c+\bar{N_c}$) representation,
only the single-trace color structure ($c=1$) would be present,
and the corresponding color factor would be smaller by a factor of $N_c$.
In this case the $U(1)$ gauge boson will {\it not} decouple from
the partial amplitude, so one should only sum over $SU(N_c)$ indices
when color-summing the cross-section.)
In each case the massless spin-$J$ particle is taken to have two
helicity states: gauge bosons, Weyl fermions, and complex scalars.

It is very convenient to organize the spin-$J$ partial amplitudes
in a supersymmetry-inspired fashion,
$$
\eqalign{
A_n^{[0]} &= A_{n}^{\rm scalar}\, , \cr
A_n^{[1/2]} &= A_{n}^{\rm fermion}
\ =\ -A_{n}^{\rm scalar} + A_{n}^{N=1 \rm \ susy}\, , \cr
A_n^{[1]} & = A_{n}^{\rm gluon}
 = A_{n}^{\rm scalar}
  - 4 A_{n}^{N=1 \rm \ susy} + A_{n}^{N=4 \rm \ susy} \, ,\cr }
\eqn\TotalAmp
$$
where the $N=1$ supersymmetry label refers to the contribution
of a chiral multiplet consisting of a complex scalar and a Weyl
fermion, while the $N=4$ supersymmetry label refers to a vector
multiplet consisting of three complex scalars, four Weyl fermions and
a single gluon.
We have assumed the use of a supersymmetry preserving regulator
[\use\Siegel,\use\StringBased,\use\Tasi,\use\KST].

The utility of the decomposition~(\use\TotalAmp) stems from the
simplicity of $N=4$ and $N=1$ supersymmetric calculations, in any
approach where supersymmetric cancellations are made manifest in each
diagram, such as string-based~[\use\StringBased,\use\Mapping] or
superspace approaches~[\use\Superspace].  The string-based approach is
related to use of the background field gauge effective action
[\use\Background] together with a second-order form for the fermion loop,
summarized by the determinants [\use\Mapping],
$$
\eqalign{
{\mit \Gamma}_{\rm scalar}[A] & = \ln \det{}^{-1} [D^2] \, , \cr
{\mit \Gamma}_{\rm fermion}[A] & = {1\over 2}
\ln\det [\; \slash \hskip - .26 cm D ]=
{1\over 2}\ln\det{}^{1/2}
[D^2 -  {g\over 2} \sigma_{\mu\nu} F^{\mu\nu}] \, , \cr
{\mit \Gamma}_{\rm gluon}[A] & = \ln
\det{}^{-1/2} [D^2 \eta_{\alpha\beta}
- g (\Sigma_{\mu\nu})_{\alpha\beta} F^{\mu\nu}] + \ln \det [D^2] \, , \cr}
\eqn\BkgdDets
$$
where $D$ is the covariant derivative, the particle labels refer to
the states circulating in the loop, ${1\over2}\sigma_{\mu\nu}$
($\Sigma_{\mu\nu}$) are the spin-${1\over2}$ (spin-1) Lorentz
generators, and we have used the fact that the contribution of a Weyl
fermion in a non-chiral theory is half that of a Dirac fermion.
(Note that the gluon determinant contains Lorentz indices and the fermion
determinant spinor indices.)  The
similarity of these determinants to each other suggests that there may
be significant overlap between the calculation of a gluon loop and the
calculation of a fermion loop.  Indeed from~(\use\BkgdDets) one can
see that for the $N=1$ chiral matter multiplet contribution (scalar
plus fermion) the pure $D^2$ contributions cancel, which means that
the $m$-point loop integrands will contain at most $m-2$ powers of the
loop momentum in the numerator, or two fewer than that for an
individual scalar, fermion or gluon contribution.  For the $N=4$
contribution, as a result of further cancellations (which are manifest
in superspace or string-based approaches), $m$-point loop integrands
contain at most $m-4$ powers of the loop momentum.  A reduction of two
powers of the loop momentum in each diagram permits one to apply a
result, described in section 5, which dictates that the cuts
completely specify the amplitude [\use\SusyOne]; thus, loop diagrams
can be bypassed in favor of simpler cut calculations for the $N=1$ and
$N=4$ supersymmetric contributions.  In a conventional formalism using
ordinary gauges and fermion Feynman rules, one would not find the
supersymmetric simplifications until all diagrams are summed, and it
would not be clear how to apply this result.

The relative simplicity of $N=4$ loop amplitudes was first observed by
Green, Schwarz and Brink in their calculation of the four-gluon
amplitude as the low-energy limit of a superstring [\use\GSB]; the
same simplicity was found to extend to five-gluon amplitudes (for both
$N=4$ and $N=1$ components) in ref.~[\use\FiveGluon].  The
supersymmetric decomposition can also reveal structure in electroweak
amplitudes that would otherwise remain hidden
[\use\Morgan,\use\WeakInt].

\vskip .2 cm
\noindent
{\bf 3. A Subleading from Leading Color Formula.}

\vskip .1 cm

It turns out that the subleading-color partial amplitudes associated
with adjoint representation states appearing
in eq.~(\use\ColorDecomposition) can
be obtained from the leading color partial amplitudes
by summing over an appropriate set of permutations given by
$$
 A_{n;c}(1,2,\ldots,c-1;c,c+1,\ldots,n)\ =\
 (-1)^{c-1} \sum_{\sigma\in COP\{\alpha\}\{\beta\}} A_{n;1}(\sigma)
\eqn\sublanswer
$$
where $\alpha_i \in \{\alpha\} \equiv \{c-1,c-2,\ldots,2,1\}$,
$\beta_i \in \{\beta\} \equiv \{c,c+1,\ldots,n-1,n\}$, and
$COP\{\alpha\}\{\beta\}$ is the set of all permutations of
$\{1,2,\ldots,n\}$ with $n$ held fixed that preserve the cyclic
ordering of the $\alpha_i$ within $\{\alpha\}$ and of the $\beta_i$
within $\{\beta\}$, while allowing for all possible relative orderings
of the $\alpha_i$ with respect to the $\beta_i$.  Note that the
ordering of the first sets of indices is reversed with respect to the
second.
This formula may be easily derived
using string-based rules [\use\SusyFour], but can also be derived in
field theory using the color-ordered Feynman rules described in
refs.~[\use\ManganoReview,\use\Tasi].
This formula eliminates the need to do
a separate calculation for the subleading-color parts of an $n$-gluon
amplitude; we therefore need only the leading-color partial amplitudes.

\vskip .2 cm
\noindent
{\bf 4. Collinear Limits Constraint.}

\vskip .1 cm

Consider first the $n$-point tree-level partial amplitude
$A_n(1,2,\ldots,n)$ with an arbitrary helicity configuration.
The external legs may be fermions or gluons.
There is an implicit color ordering of the vertices $1,2,\ldots,n$,
so that collinear singularities arise only from neighboring
legs $a$ and $b$
becoming collinear~[\use\TreeCollinear,\use\ManganoReview].  These
singularities have the form
$$
A_{n}^{\rm tree}\ \mathop{\longrightarrow}^{a \parallel b}\
\sum_{\lambda=\pm}
 \Split^{\rm tree}_{-\lambda}(a^{\lambda_a},b^{\lambda_b})\,
      A_{n-1}^{\rm tree}(\ldots(a+b)^\lambda\ldots)\ ,
\eqn\treesplit
$$
where the non-vanishing splitting amplitudes diverge as
$1/\sqrt{s_{ab}}$ in the collinear limit $s_{ab}=(k_a+k_b)^2\rightarrow0$.
In the collinear limit $k_a = z\,P$, $k_b = (1-z)\,P$, where $P$ is
the sum of the collinear momenta; $\lambda$ is the helicity of the
intermediate state with momentum $P$.
The tree splitting amplitudes
$\Split^{\rm tree}_{-\lambda}(a^{\lambda_a},b^{\lambda_b})$
may be found in
refs.~[\use\ParkeTaylor,\use\ManganoParke,\use\RecursiveA,\use\ManganoReview].

The collinear limits of the (color-ordered) one-loop partial
amplitudes have the form
$$
\eqalign{
A_{n;1}^{\rm loop}\ \mathop{\longrightarrow}^{a \parallel b}\
\sum_{\lambda=\pm}  \biggl(
  \Split^{\rm tree}_{-\lambda}(a^{\lambda_a},b^{\lambda_b})\,
&
      A_{n-1;1}^{\rm loop}(\ldots(a+b)^\lambda\ldots)
\cr
&  +\Split^{\rm loop}_{-\lambda}(a^{\lambda_a},b^{\lambda_b})\,
      A_{n-1}^{\rm tree}(\ldots(a+b)^\lambda\ldots) \biggr) ,
\cr}
\eqn\loopsplit
$$
which is schematically depicted in \fig\CollinearFigure .
The splitting amplitudes
$\Split^{\rm tree}_{-\lambda}(a^{\lambda_a},b^{\lambda_b})$ and
$\Split^{\rm loop}_{-\lambda}(a^{\lambda_a},b^{\lambda_b})$
are universal: they depend only on the two legs becoming
collinear, and not upon the specific amplitude under consideration.
This universal behavior is expected to hold for all one-loop amplitudes,
with external (massless) fermions as well as gluons;
all one-loop amplitudes that we have inspected do indeed obey
eq.~(\use\loopsplit).  (A similar equation is expected to govern the
limit of one-loop partial amplitudes as one external
gluon momentum becomes soft.)
The explicit $\Split^{\rm loop}_{-\lambda}(a^{\lambda_a},b^{\lambda_b})$
have been determined from the known four- and five-point
one-loop amplitudes~[\use\FiveGluon,\use\Fermion], and are collected in
appendix~II of ref.~[\use\SusyFour].
An outline of a direct proof of the universality of the splitting
amplitudes for the scalar-loop contributions to amplitudes with
external gluons was presented in ref.~[\use\AllPlus]; a more general
discussion will be presented elsewhere [\use\Splitting].

\vskip .1 cm
\centerline{\epsfxsize 3.7 truein \epsfbox{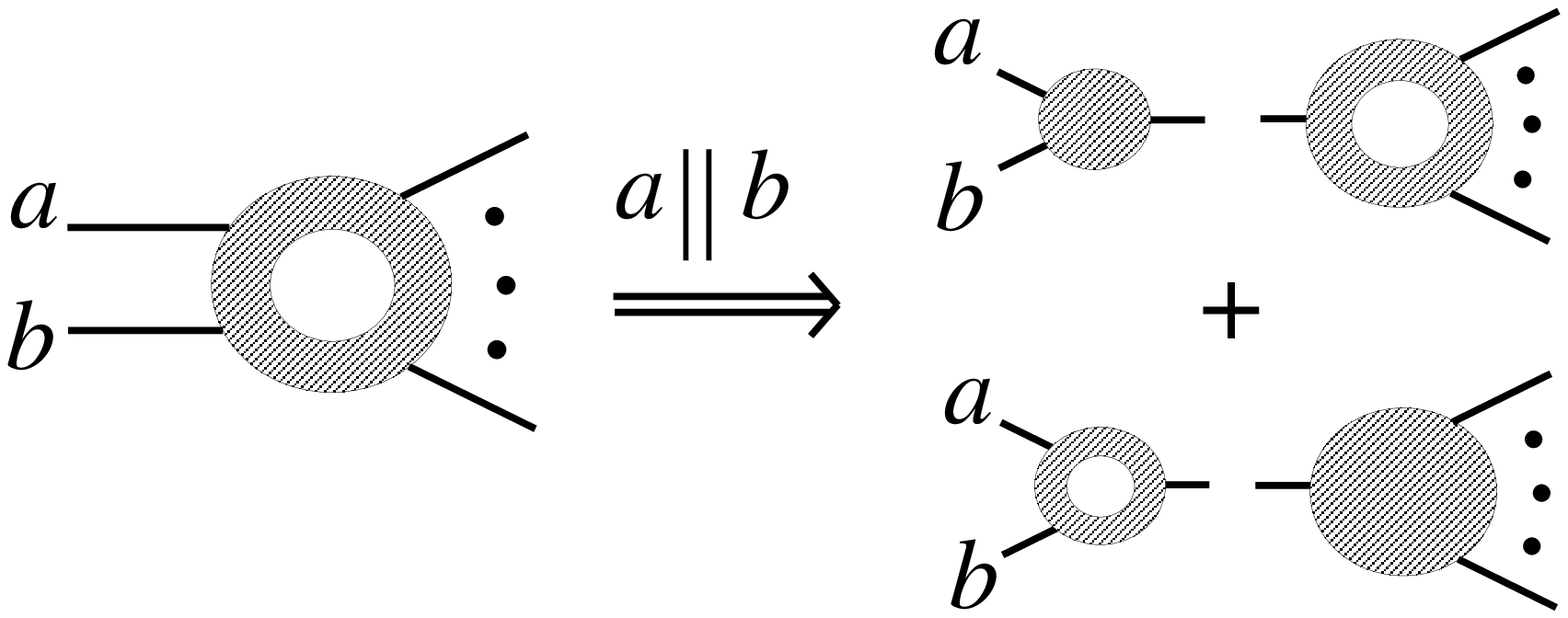}}
\nobreak
\vskip -.02 cm\nobreak
{ \narrower\smallskip \ninerm
\noindent{\ninebf Fig.~\CollinearFigure:} A schematic representation
of the behavior of one-loop amplitudes as the momenta of two legs become
collinear.
\smallskip}

\vskip .1 cm

The collinear behavior places tight constraints on the possible form
of one-loop amplitudes and allows one to construct ans\"atze for
higher-point amplitudes.  To do this one writes down a general form
for a higher-point amplitude containing arbitrary coefficients which
may then be fixed by demanding that the expression have the correct
collinear behavior.  In this way a `collinear bootstrap' can be
constructed in certain cases to an arbitrary number of external legs
[\use\AllPlus].

However, functions lacking singular behavior in any
collinear limit may appear in amplitudes, and would thus
be omitted in such a construction. The simplest
non-trivial example of such a function is the five-point function
$$
{\varepsilon(1,2,3,4)
\over \spa1.2 \spa2.3 \spa3.4 \spa4.5 \spa5.1}
\ ,
\eqn\fivepointambiguity
$$
since the contracted antisymmetric tensor
$\varepsilon(1,2,3,4) \equiv  4i \varepsilon_{\mu\nu\rho\sigma}
 k_1^\mu k_2^\nu k_3^\rho k_4^\sigma$
vanishes when any two of the five vectors
$k_i$ become collinear ($\sum_{i=1}^5 k_i = 0$).
Another example is the six-point function
$$
\sum_{P(1,\ldots,5)}
{ \ln(-s_{12}) + \ln(-s_{23}) + \cdots + \ln(-s_{61})
\over \spa1.2 \spa2.3 \spa3.4 \spa4.5 \spa5.6 \spa6.1 }  \, ,
\eqn\sixpointambiguity
$$
where the
summation is over all 120 permutations of legs 1 through 5 and
$s_{i, i+1} = (k_i + k_{i+1})^2$.
Without additional information, functions
such as~(\use\fivepointambiguity) and (\use\sixpointambiguity)
represent additive ambiguities in the collinear bootstrap.

\vskip .2 cm
\noindent
{\bf 5. Unitarity Constraint.}
\vskip .1 cm

The Cutkosky cutting rules~[\use\Cutting] provide a relatively simple
way to fix the absorptive (cut) parts of amplitudes.
These cut amplitudes are generally much simpler to
evaluate than the full amplitude.  We calculate cuts in
terms of the imaginary parts of one-loop integrals that would have
been encountered in a direct calculation.  This makes it
straightforward to write down an analytic expression with the correct
cuts in all channels, thus avoiding the need to do a dispersion
integral to reconstruct the full amplitudes.

In order to evaluate the cuts, consider the amplitude, not in a
physical kinematic configuration, but in a region where exactly one of
the momentum invariants is taken to be positive (time-like), and the
rest are negative (space-like).  In this way cuts are isolated in a
single momentum channel.  The Cutkosky rules are applied at the
amplitude level, rather than at the diagram level.  That is, write the
sum of all cut diagrams as the sum of all tree diagrams on one side of
the cut, multiplied by the sum of all tree diagrams on the other side
of the cut.  Thus the cut in the one-loop amplitude is given by the
integral over a two-body phase-space of the product of two tree
amplitudes, which is then summed over each intermediate helicity
configuration that contributes.  By summing over the various cuts one
can construct an expression which has all the correct cuts in all
channels.

Unitarity determines the cuts uniquely --- and hence the dilogarithms
and logarithms --- but does not directly provide any information about
polynomial terms in the amplitude.  (By `polynomial terms' we actually
mean any cut-free function of the kinematic invariants and spinor
products, that is any rational function of these variables.)

The supersymmetric case is however special: a knowledge of the cuts
completely determines the amplitude.  The key to this result is the
property discussed in section~2, that in a supersymmetric theory the
loop-momentum polynomials encountered in every string-based or
superspace diagram have a degree that is at least two less than the
purely gluonic case, namely $m-2$ for an $m$-point integral.  This
means that only a restricted set of integrals (after reduction [\use\PV]
to boxes, triangles, and bubbles [\use\MVNV,\use\Integrals]) can appear in an
explicit calculation of supersymmetric amplitudes.  One can
show that there is no linear combination of integrals
(with coefficients rational functions of the momentum invariants)
in this restricted set which is free of cuts, that is which yields
a `polynomial term' in the sense used above.
  Further details of this procedure are
given in refs.~[\use\SusyFour,\use\SusyOne].

In general, for theories other than supersymmetric ones, the cuts may
not uniquely determine the full amplitude.  As a simple example, the
five-point helicity amplitudes $A_{5;1}(1^-,2^+,\ldots,5^+)$ and
$A_{5;1}(1^+,2^+,\ldots,5^+)$ each have no nontrivial cuts but are not equal.
In such cases
the collinear limits provide restrictions on the form of rational
functions that may appear in the amplitudes~[\use\AllPlus].

\vskip .2 cm
\noindent {\bf 6. Obtaining New Amplitudes from Known Amplitudes.}
\vskip .1 cm

Using string-based methods all one-loop five-gluon helicity amplitudes
[\use\FiveGluon] have been computed. Besides the intrinsic value of
these amplitudes to the computation of next-to-leading order
corrections to the three-jet cross-section at hadron colliders,
these amplitudes are also useful as a starting point to generate
further amplitudes.  One can use the collinear bootstrap
discussed in section 4 to obtain amplitudes with a larger number of external
legs.  Supersymmetry identities can also be used to generate
contributions to amplitudes with external fermions
[\use\Susy,\use\SusyTree,\use\KST].  We now present examples of
amplitudes and point out how they are used to generate further
amplitudes.

One amplitude that we present is for
 the all-plus helicity configuration
$$
\eqalign{
A_{5;1} & (1^+,2^+,3^+,4^+,5^+)  = \cr
& \hskip 1 cm
{iN_p\over 192\pi^2}\,
  {  s_{12}s_{23} + s_{23}s_{34} + s_{34}s_{45} + s_{45}s_{51} +
     s_{51}s_{12}\ +\ \pol(1,2,3,4)
   \over \spa1.2 \spa2.3 \spa3.4 \spa4.5 \spa5.1 }\ ,  \cr }
\eqn\AllPlusFive
$$
where
$\pol(i,j,m,n) = 4i\varepsilon_{\mu\nu\rho\sigma}
        k_i^\mu k_j^\nu k_m^\rho k_n^\sigma
    \ =\ \spb{i}.{j}\spa{j}.{m}\spb{m}.{n}\spa{n}.{i}
       - \spa{i}.{j}\spb{j}.{m}\spa{m}.{n}\spb{n}.{i}
$, and $N_p$ is the number of color-weighted bosonic states
minus fermionic states circulating in the loop;
for QCD with $n_f$ quarks, $N_p = 2(1-n_f/N_c)$ with the number of
colors $N_c=3$. This amplitude forms the basis for a collinear
bootstrap to larger numbers of external legs, discussed in
the next section.

Another example which can be extended to an arbitrary number of
external legs is the set of $N=4$ supersymmetric five-gluon amplitudes
$$
\hskip -.4 cm
A_{5;1}^{N=4\ \rm susy} = \cg A_{5}^{\rm tree} \biggl[
\sum_{i=1}^{5} -{ 1 \over \eps^2 } \Bigl(
{ \mu^2  \over -s_{i,i+1} } \Bigr)^{\eps}
           +\sum_{i=1}^5 \ln\L{-s_{i,i+1} \over-s_{i+1,i+2} }\R\,
               \ln\L{-s_{i+2,i+3}\over-s_{i-2,i-1}}\R+{5\over6}\pi^2 \biggr]
\eqn\SusyFourAmpl
$$
where $A_5^{\rm tree}$ is the tree amplitude for the same (MHV) helicity
configuration, given by eq.~(\use\PT) for $n=5$.
The overall constant is
$$
\cg = {(4 \pi)^\eps \over 16 \pi^2 }
{\Gamma(1+\eps)\Gamma^2(1-\eps)\over\Gamma(1-2\eps)}\ .
\eqn\Prefactor
$$
In this case, the cuts turn out to provide a more powerful tool for
extending~(\use\SusyFourAmpl) to an arbitrary number of legs,
than do the collinear limits,
because in the $N=4$ case the cuts uniquely determine the
amplitude [\use\SusyFour].

As an example of how supersymmetry can be used to aid in the
calculation of amplitudes with external fermions, consider
$A_{5;1}(1_{\bar{q}}^-,2_q^+,3^+,4^+,5^-)$, where the first two legs
are fermionic and the remaining ones gluonic.  In QCD this
partial amplitude is associated with the leading color factor
$(T^{a_3} T^{a_4} T^{a_5})_{i_2}{}^{\bar i_1}$.  For the calculation
of five-point leading-color amplitudes with external
fermions [\use\Fermion], we
used a field theory approach containing a number of improvements
motivated by superstring theory, including better gauge choices,
supersymmetry decompositions, and an improved decomposition into gauge
invariant pieces. A more recent development has been to make use of
the observation that the cuts completely determine large parts of the
amplitudes.

To illustrate the use of supersymmetry identities, first consider
the contribution of an $N=1$
chiral multiplet (one complex scalar and one Weyl fermion) to the known
five-gluon amplitude with the same helicity configuration given by
[\use\FiveGluon]
$$
\eqalign{
A^{N=1\ \rm susy} & (1^-, 2^+, 3^+, 4^+, 5^-) = \cr
& \cg A_5^{\rm tree}(1^-, 2^+, 3^+, 4^+, 5^-) \biggl\{
  {1\over2\e(1-2\e)} \L \L{\mu^2\over-s_{12}}\R^\e
                +  \L{\mu^2\over-s_{45}}\R^\e \R  \cr
\null & \hskip 3 cm
  - {1\over2 s_{51}}
     \Bigl( \tr[5123] + \tr[5134] \Bigr)
    {\Ll_0\L {-s_{12}\over -s_{45}}\R \over s_{45}} \biggr\} \, ,   \cr}
\eqn\SusyOneAmpl
$$
where
$$
\eqalign{
\tr[a_1 a_2 \cdots a_{2 m}] & \equiv
\tr_+[\ksl_{a_1}\ksl_{a_2} \cdots \ksl_{2m}] \equiv
{1\over 2} \tr[(1+\gamma_5)\ksl_{a_1}\ksl_{a_2} \cdots \ksl_{2m}] \cr
& =
\spb{a_1}.{a_2}\spa{a_2}. {a_3} \cdots
\spb{a_{2m-1}}.{a_{2m}}\spa{a_{2m}}.{a_1} \, , \cr}
\anoneqn
$$
and
$$
\Atree_5( 1^{-},2^{+},3^{+},4^{+},5^{-})
= i {{\spa1.5}^4\over \spa1.2\spa2.3\spa3.4\spa4.5\spa5.1} \, .
\anoneqn
$$
The poles in $s_{ij}$ are fictitious and cancel against factors in the
numerators and in $A_5^{\rm \tree}$.  We have defined an auxiliary
set of functions,
$$
\eqalign{
&  \Ll_0(r) = {\ln(r)\over 1-r}\,,\hskip 10mm
  \Ll_1(r) = {\ln(r)+1-r\over (1-r)^2}\,,\hskip 10mm
  \Ll_2(r) = {\ln(r)-\hf(r-1/r)\over (1-r)^3}\,,\cr
&\Ls_1(r_1,r_2) = {1\over (1-r_1-r_2)^2}
  \LB \Li_2(1-r_1) + \Li_2(1-r_2) + \ln r_1\,\ln r_2 - {\pi^2\over6}\RP\cr
 &\hskip 30mm\LP\vphantom{{\pi^2\over6}}
   + (1-r_1-r_2) \L \Ll_0(r_1) + \Ll_0(r_2)\R \RB\ ,\cr
}\eqn\Lsdef
$$
and $\Li_2(x) = -\int_0^x dz \ln(1-z)/z$ is the
dilogarithm [\use\Lewin].  In eq.~(\use\SusyOneAmpl),
as for all subsequent amplitudes,
no ultraviolet subtraction has been performed;
in modified minimal subtraction the quantity $-3 A^{\rm tree} \cg/2\eps$
should be subtracted from $A^{N=1\ \rm susy}$.
Note that there are no polynomial terms in~(\use\SusyOneAmpl)
that are independent of the terms with cuts, in accordance
with the result discussed in section 5, that for a supersymmetric
amplitude the cuts uniquely specify the amplitude.

Now consider the corresponding
amplitude with two external quarks and three gluons.
In order to make the supersymmetry relations more apparent
decompose the loop amplitude in terms of
supersymmetric and non-supersymmetric pieces
$$
\eqalign{
  A_{n;1}(1_{\bar{q}},2_q;3,\ldots,n) &
   =  A^{N=4\ \rm susy} - 3 A^{N=1\ \rm susy}\
   -\ \left(1+{1\over N^2}\right) A^\Ram \cr
\null & \hskip 2 cm
    \ +\ \left(1+{n_s\over N}-{n_f\over N}\right) A^s
    \ +\ \left(1-{n_f\over N}\right) A^f \ , \cr}
\eqn\Anonedecomp
$$
where $A^{N=4\ \rm susy}$ and $A^{N=1\ \rm susy}$ are the
contributions of an $N=4$ vector super-multiplet and
an $N=1$ chiral matter multiplet.  This decomposition,
although more complicated, is analogous to the one given in
eq.~(\use\TotalAmp) for purely external gluons.
The $N=4$ and $N=1$ contributions are given from the corresponding five-gluon
amplitudes in eqs.~(\use\SusyFourAmpl) and (\use\SusyOneAmpl)
by the supersymmetry Ward identity [\use\Susy]
$$
A_{5;1}^{\rm susy} (1_{\bar{q}}^-,2_q^+,3^+,4^+,5^-)
= -{\spa2.5 \over \spa1.5} A_{5;1}^{\rm susy} (1^-,2^+,3^+,4^+,5^-)\, ,
\anoneqn
$$
so there is actually no need to recalculate the supersymmetric parts
of the amplitudes;
alternatively one can use the supersymmetry identity as a non-trivial check
on results, which we have performed for this amplitude.
The other components of this helicity amplitude are
$$
\hskip -.4 cm
\eqalign{
A^\Ram\ &=\ \cg A^{\rm tree} \Biggl\{
 - {1\over\e^2} \L{\mu^2\over-s_{12}}\R^\e
  - {3\over2\e} \L{\mu^2\over-s_{12}}\R^\e - {7\over2}
 +   {\tr[1 5 4 3] \tr[1 5 3 4 5 2] \over 2 s_{51}^2 s_{25}}
    {\Ll_1\L {-s_{12}\over -s_{45}}\R \over s_{45}^2} \cr
&\quad
 - {s_{14} \tr[1 4 5 2] \over s_{25}}
    {\Ls_1\L {-s_{45}\over -s_{23}},\,{-s_{51}\over -s_{23}}\R
         \over s_{23}^2}
 - {s_{13} \tr[1 3 5 2] \over s_{25}}
    {\Ls_1\L {-s_{12}\over -s_{45}},\,{-s_{23}\over -s_{45}}\R
         \over s_{45}^2}
+ {\tr[1 4 5 2] \over s_{25}}
    {\Ll_0\L {-s_{23}\over -s_{51}}\R \over s_{51}}
\cr
&\quad
- s_{12}
    {\Ll_0\L {-s_{23}\over -s_{45}}\R \over s_{45}}
 + {\tr[1 5 3 2] \over s_{51}}
    {\Ll_0\L {-s_{12}\over -s_{45}}\R \over s_{45}}
 +  {\tr[1 5 4 2] \tr[3 4 5 2] \over  2s_{51}^2 s_{25} s_{45}} \Biggr\} \ , \cr
A^s\ &=\ -{\cg\over 3} A^{\rm tree} \Biggl\{
 - 2{\tr^2[1 5 4 3] \tr[3 4 5 2] \over s_{51}^2 s_{25}}
    {\Ll_2\L{-s_{45}\over -s_{12}}\R \over s_{12}^3 }
 + 3{\tr[1 5 4 3] \tr[3 4 5 2] \over s_{51} s_{25}}
    {  \Ll_1\L{-s_{45}\over -s_{12}}\R \over s_{12}^2 }   \cr
&\quad\hskip 2 cm
 - 2 {\tr[3 4 5 2] \over s_{12} s_{25} }
 + {\tr[1 3 2 4 5 2] \over s_{12} s_{51} s_{25}}
 + {s_{14} \tr[1 5 4 2] \tr[3 4 5 2] \over s_{12} s_{51}^2 s_{25} s_{45}}
            \Biggr\} \ ,\cr
A^f\ &=\ \cg A^{\rm tree} {\tr[3 4 5 2] \over s_{25}}
   { \Ll_0\L{-s_{12}\over -s_{45}}\R \over s_{45} } \ .  \cr }
\eqn\ffppmfunctionsb
$$
(Again no ultraviolet subtraction has been performed and we have
used a supersymmetry preserving regulator
[\use\Siegel,\use\StringBased,\use\Tasi,\use\KST].)
The tree amplitude for this helicity configuration is
$$
A_5^{\rm tree}(1_{\bar{q}}^-,2_q^+,3^+,4^+,5^-) =
 -i \, {{\spa1.5}^3\spa2.5 \over \spa1.2\spa2.3\spa3.4\spa4.5\spa5.1}\, .
\anoneqn
$$
Terms containing logarithms and dilogorithms may be computed most
efficiently via the Cutkosky rules.  The polynomial terms,
however, are significantly more difficult to compute.
Even though expression (\use\ffppmfunctionsb) is complicated,
it should be possible to extend this result to at least
one more leg via the `collinear-unitarity bootstrap' discussed in
sections 4 and 5.

\vskip .2 cm
\noindent {\bf 7. The All-Plus Helicity Amplitudes.}

\vskip .1 cm

The structure of $A_{n;1}(1^+,2^+, \cdots n^+)$ is particularly
simple, making it an ideal first candidate for finding an all-$n$
expression.  The all-plus helicity structure is cyclicly symmetric,
and no logarithms or other functions containing branch cuts can
appear.  This can be seen by considering the cutting rules: the cut in
a given channel is given by a phase space integral of the product of
the two tree amplitudes obtained from cutting.  One of these tree
amplitudes will vanish for all assignments of helicities on the cut
internal legs since $A_{n}^{\rm tree} (1^\pm,2^+,3^+,\ldots,n^+)=0$,
so that all cuts vanish.
For the same reason, consideration of factorization on poles in the
sum of three or more momenta shows that the all-plus
helicity loop amplitude does not contain such multi-particle poles.

The starting point in constructing an $n$-point expression is the
previously calculated~[\use\FiveGluon]
five-point one-loop helicity amplitude~(\use\AllPlusFive).
Using eq.~(\use\loopsplit), the explicit form of the tree splitting
amplitudes
[\use\TreeCollinear,\use\ManganoReview],
$A^{\rm tree}_n (1^\pm, 2^+, \cdots, n^+) = 0$, and experimenting
at small $n$,
higher point amplitudes can be constructed by writing down general forms
with only two particle-poles,
and requiring that they have the correct collinear limits.
Doing so, one may obtain the all $n$ ansatz~[\use\BDKconf,\use\AllPlus],
$$\eqalign{
 A_{n;1}(1^+,2^+,\ldots,n^+)\ =\ -{i N_p \over 96\pi^2}\,
\sum_{1\leq i_1 < i_2 < i_3 < i_4 \leq n}
{ {\rm tr}_-[i_1 i_2 i_3 i_4]
\over \spa1.2 \spa2.3 \cdots \spa{n}.1 }\ ,
}\eqn\allnplus
$$
where $\tr_-[i_1 i_2 i_3 i_4] =
{1\over 2}\tr_-[(1-\gamma_5) \ksl_{i_1} \ksl_{i_2} \ksl_{i_3} \ksl_{i_4}]$.
This has been confirmed by a recursive calculation~[\use\MahlonB].

In massless QED, through use of recursion relations,
Mahlon has demonstrated that
the one-loop $n$-photon helicity amplitudes
$A_n(\gamma_1^\pm,\gamma_2^+, \cdots,\gamma_n^+)$ vanish for $n>4$
[\use\MahlonA].
It is easy to argue
that the collinear limits are consistent with this result, and that
many more ``mixed'' photon-gluon amplitudes should also vanish.
Charge conjugation invariance
implies that photon amplitudes with an odd number of
legs vanish.  This also implies that the amplitude with
three photons and two gluons $A_{5;1}(\gamma_1, \gamma_2,
\gamma_3, g_4, g_5) =0$, since this amplitude is proportional to the
corresponding five-photon amplitude.  Using the collinear
behavior~(\use\loopsplit) leads one to suspect that the six-point
helicity amplitudes $A_{6;1}(\gamma_1^\pm , \gamma_2^+, \gamma_3^+,
g_4^+, g_5^+, g_6^+)$ and $A_{6;1}(\gamma_1^+ , \gamma_2^+,
\gamma_3^+, g_4^+, g_5^+, g_6^\pm)$ may vanish since both terms on the
right-hand-side of the collinear limits formula (\use\loopsplit)
vanish. Continuing recursively in this way, one may surmise that
all other one-loop cut-free amplitudes with three photons and
additional gluons vanish.
Photon-gluon amplitudes which contain cuts ---
those with two or more helicities of opposite sign from the rest
--- need not vanish by the same argument, due to the
existence of functions such as~(\use\sixpointambiguity),
which have cuts but are nonsingular in every collinear limit.

To verify that the all-plus helicity amplitude with three photons and
$(n-3)$ gluons vanishes, one can convert three of the gluons in expression
(\allnplus) into photons.  Amplitudes with $r$ external photons and
$(n-r)$ gluons have a color decomposition similar to that of the
pure-gluon amplitudes, except that charge matrices are set to unity
for the photon legs.  This implies that partial amplitudes with photons
are given by sums over the permutations of the $n$-gluon partial
amplitude that leave the color trace factor for the $(n-r)$
``true'' gluons invariant.
Explicitly performing the permutation sum for three external photons,
one finds that the amplitude vanishes,
$$
A^{\rm loop}_{n>4}(\gamma_1^+, \gamma_2^+, \gamma_3^+, g_4^+, \ldots, g_n^+)
 =0 \, ,
\eqn\ThreePVanish
$$
in agreement with the collinear limits argument.
Since amplitudes with even more photon legs are
obtained by further sums over permutations of legs, the all-plus
helicity amplitudes with three or more photon legs vanish (for $n>4$)
in agreement with the expectation from the collinear limits.
Equation~(\use\ThreePVanish) has been shown to hold also in the case
where one of the photon legs is of opposite helicity [\use\DP].
There is numerical evidence that it is true as well in the remaining
cut-free case, when one of the gluon legs is of opposite
helicity~[\use\MahlonP].

\vskip .2 cm
\noindent {\bf 8. Amplitudes with Two Negative Helicities.}

\vskip .1 cm

In this section amplitudes with two negative helicities
(MHV amplitudes) are considered, as
these contribute to next-to-leading order multi-jet cross-sections.
As discussed in section 2, it is useful to decompose a QCD amplitude
into supersymmetry-based pieces.  In particular, the $n$-gluon amplitude
may be split up as in eq.~(\TotalAmp).
As the $N=4$ supersymmetric amplitude is the simplest piece of an
$n$-gluon QCD amplitude we discuss it first.  One can again
use the collinear limits to construct an all-$n$ $N=4$ MHV ansatz
starting from the five-point amplitude~(\use\SusyFourAmpl).
Such a construction is similar to the previous all-plus example.
It turns out, however,
that in the supersymmetric case the unitarity cuts provide
a more powerful tool for obtaining an all-$n$ formula for the amplitude,
because the cuts fix the amplitude uniquely [\use\SusyFour].

The cuts can be evaluated directly from the Cutkosky rules~[\use\Cutting],
which turn out to require only the Parke-Taylor (MHV) tree
amplitudes~(\use\PT) for their evaluation.  For example for
the cut where the two negative helicity gluons are on the same side
of the cut is
$$
\hskip -.4 cm
\eqalign{
  &\int \dlips(-\ell_1,\ell_2)
  \ A^\treemhv_{jk}(-\ell_1,m_1,\ldots,m_2,\ell_2)
  \ A^\treemhv_{(-\ell_2)\ell_1}(-\ell_2,m_2+1,\ldots,m_1-1,\ell_1), \cr}
\anoneqn
$$
where $\dlips(-\ell_1,\ell_2)$ denotes the Lorentz-invariant phase space
measure.  After replacing the cut propagators with ordinary
propagators we obtain, for any cut and any value of $n$, a single cut hexagon
integral to evaluate, which can be further reduced to a sum of four cut
scalar box integrals.
Combining the expressions for the various cuts,
one obtains a function whose cuts all match the cuts of the amplitude.
The result for the one loop $N=4$ MHV amplitudes with
an arbitrary number of external legs are
schematically depicted in \fig\SusyFourFigure\ in terms of box
integral functions.  The explicit expression, together
with a more detailed derivation, may be found in ref.~[\use\SusyFour].

\vskip -.2 cm
\centerline{\epsfxsize 4.2 truein \epsfbox{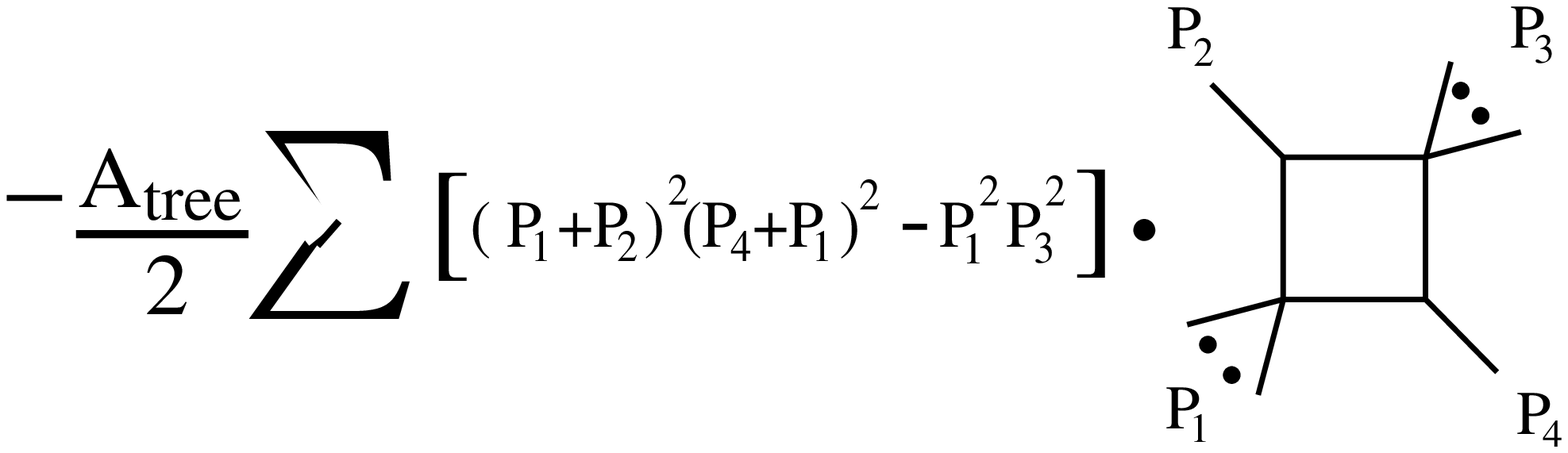}}
\nobreak
\vskip -8.5 cm\nobreak
{ \narrower\smallskip \ninerm
\noindent{\ninebf Fig.~\SusyFourFigure:} A schematic representation of
the $N=4$ supersymmetry MHV amplitudes.
\smallskip}

\vskip .1 cm

The contributions of an $N=1$ chiral multiplet to MHV amplitudes have
also been constructed via their cuts~[\SusyOne].  The explicit form
where the two negative helicity legs are nearest neighbors is given by
$$
\eqalign{
A_{n;1}^{N=1\ \rm susy}& (1^-, 2^-, 3^+, \cdots, n^+) =  \cr
& \cg {A^{\rm tree}_n} \biggl\{
 {1\over2\e(1-2\eps)}\LB \L{\mu^2\over-s_{23}}\R^\e
                         +\L{\mu^2\over-s_{n1}}\R^\e \RB \cr
& +{1\over 2  s_{12} }
   \sum_{m=2}^{n-3} {\Ll_0\L{ -\tn{m}2 \over -\tn{m+1}2} \R\over \tn{m+1}2}
( {\rm tr_+ }[\ksl_1 \ksl_2 \ksl_{m+2} \qsl_m ]
-{\rm tr_+ } [\ksl_1 \ksl_2 \qsl_m \ksl_{m+2} ]) \biggr\} \, , \cr}
\anoneqn
$$
where $q_m=\sum_{j=2}^{m+1} k_j$, and $t_2^{[m]} = q_m^2$.
As a check, we have confirmed that this
expression has all the correct collinear limits.

The $N=1$ and $N=4$ amplitudes are two parts of the supersymmetry-based
decomposition of the QCD gluon amplitude; the third piece is the
contribution of a scalar in the loop, which tends to be the most
complicated of the three.
The terms in the scalar contribution containing logarithms and
dilogarithms can of course be computed using the unitarity cuts,
but there are additional `polynomial' terms which cannot be determined
in this way; however, they should be amenable either to the `collinear
bootstrap' technique [\use\AllPlus,\use\SusyFour] or to a recursive
technique [\use\MahlonA,\use\MahlonB].

For supersymmetric theories, supersymmetry Ward identities can be
used to generate amplitudes with external fermions from the $n$-gluon
amplitudes presented above, in the same way as discussed in section 6
for five-gluon amplitudes.

\vskip .2  cm
\noindent {\bf 9. Prospects for Two Loops.}

\vskip .1 cm

As an exercise, it is not difficult to calculate the cuts in
the two-loop leading-color
amplitude $A_{4}^{\rm two-loop} (1^+, 2^+, 3^+, 4^+)$.
The four non-vanishing cuts are depicted in \fig\TwoLoopFigure. The
three-particle cuts vanish since one side of the cut will always
contain a tree which vanishes by eq.~(\use\TreeVanish), making the absorptive
parts
particularly easy to evaluate.  Another
simplifying feature is the absence of logarithms in the one-loop amplitude
$$
A_{4;1}^{\rm one-loop} (1^+, 2^+, 3^+, 4^+) =  {i\over 48 \pi^2}
{s t \over \spa1.2 \spa2.3 \spa3.4 \spa 4.1} \; ,
\anoneqn
$$
where $s = (k_1 + k_2)^2$ and $t = (k_2 + k_3)^2$ are the usual
Mandelstam variables. This means that the two-loop cuts can be
evaluated entirely in terms of one-loop integrals.
The result for the pure glue contributions
to the four-gluon leading-color partial amplitude is
$$
A_{4}^{\rm two-loop} (1^+, 2^+, 3^+, 4^+)
\sim -\cg (\mu^2)^\eps A^{\rm one-loop}_{4;1}
{2\over \eps^2} \Bigl((-s)^{-\eps} + (-t)^{-\eps}\Bigr)
+ \hbox{polynomial} \, .
\anoneqn
$$
The form of the $1/\eps^2$ terms can also be determined from the requirement
that the singularities cancel against singularities encountered
in phase-space integrals of physical cross-sections.
The cuts show that no further logarithms are present in the amplitude.
The polynomial terms however cannot be determined by unitarity,
nor can they be determined from collinear limits since there is
no lower-point amplitude from which to extrapolate.

\vskip -.2 cm
\centerline{\epsfxsize 5.5 truein \epsfbox{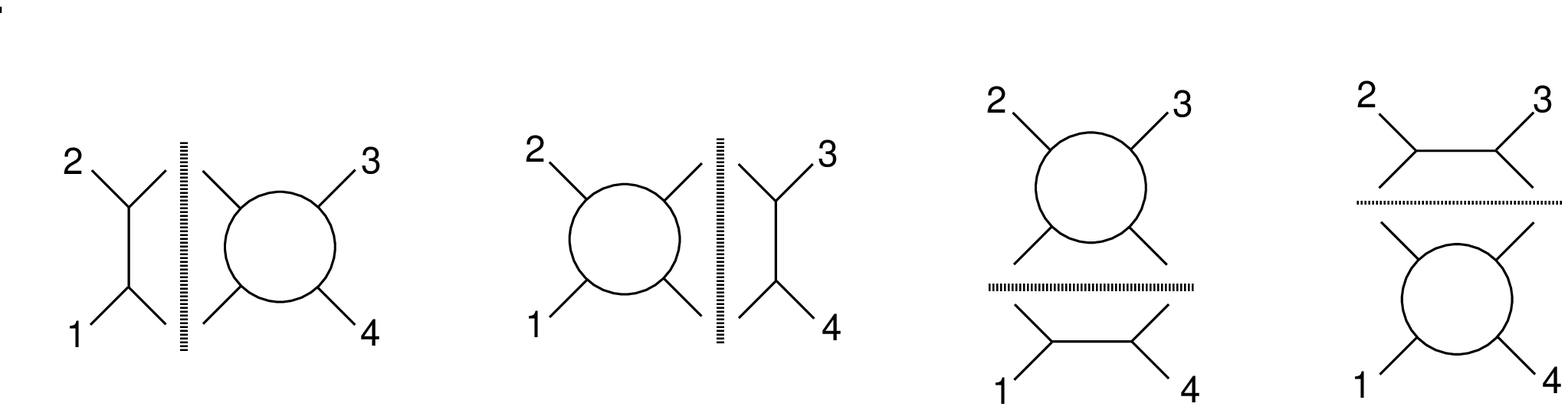}}
\nobreak
\vskip -12 cm\nobreak
{ \narrower\smallskip \ninerm
\hskip 1 cm {\ninebf Fig.~\TwoLoopFigure:} The non-vanishing cuts
in the two loop amplitude.}

\vskip .3 cm

The polynomial terms, and more generally other
two-loop amplitudes, await further advances
in calculational technology.
One must construct a two-loop integral table; and further
improvements in the formalism will also
be important.  Two possible avenues of approach are either string-based
[\use\Roland] or recursive techniques [\use\MahlonA,\use\MahlonB]. We
are hopeful that two-loop calculations will be made
possible through extensions of one-loop techniques.

\vskip .2 cm
\noindent{\bf 10. Summary and Conclusions.}
\vskip .1 cm

In this talk we presented techniques for obtaining new one-loop
amplitudes from known ones.  From the collinear (or soft) limits
[\use\TreeCollinear] one constructs higher amplitudes by demanding the
correct collinear behavior [\use\ParkeTaylor,\use\AllPlus].  From the
Cutkosky rules [\use\Cutting] one can use known tree amplitudes to
generate terms in one-loop amplitudes containing cuts.  In certain
cases, such as supersymmetric amplitudes, the cuts uniquely determine
the amplitude [\use\SusyFour,\use\SusyOne]. Finally, supersymmetry
Ward identities [\use\Susy,\use\SusyTree] can be used to obtain
supersymmetric parts of one-loop amplitudes [\use\KST] with external
fermions, using known gluon amplitudes.

Using these techniques, we have obtained a variety of exact one-loop
results for amplitudes.  In particular, we have constructed, using
unitarity constraints, one-loop supersymmetric $n$-gluon amplitudes
with maximal helicity violation.  For these amplitudes the
supersymmetry Ward identities can be immediately used to obtain
amplitudes with two external fermions and additional gluons.  We also
presented a five-point QCD example which illustrates the use of
supersymmetry to obtain parts of amplitudes with external fermions
from gluon amplitudes.  The supersymmetric amplitudes may be
interpreted as two of three terms which make up a QCD gluonic
amplitude, following the organization suggested by the string-based
method [\use\FiveGluon,\use\Tasi,\use\WeakInt].  The third piece is in
general more complicated and cannot be obtained by unitarity alone.
For example, in the all-plus helicity amplitudes the cuts are trivial
(they all vanish) and it is the use of collinear limits that allowed
us to give an ansatz for the amplitude~[\use\BDKconf,\use\AllPlus],
which was later proven correct [\use\MahlonB].  All five-gluon
amplitudes have been computed [\use\FiveGluon,\use\Fermion], and thus
provide a starting point for ans\"atze for the polynomial terms in
higher-point amplitudes, relying on the constraints imposed by the
collinear limits.  (Ambiguities in the collinear bootstrap make it
problematic to start from four-point amplitudes.)  In general, the
restrictions imposed by collinear behavior and unitarity complement
each other.

In summary, we have constructed certain classes of one-loop amplitudes
with an arbitrary number of external gluons through the reliance on
the dual constraints of unitarity and collinear behavior.  We expect
that this method will generate further fixed-$n$ and all-$n$
amplitudes.

\vskip .2 cm
We thank G. Chalmers for collaboration on work described here.  This
work was supported by the US Department of Energy Grants
DE-FG03-91ER40662 (ZB and DCD) and DE-AC03-76SF00515 (LD), by the NSF
under grant PHY-9218990 (DCD), by the Direction des Sciences de la
Mati\`ere of the Commissariat \`a l'Energie Atomique of France (DAK),
by the Alfred P. Sloan Foundation Grant APBR-3222 (ZB) and by NATO
Collaborative Research Grants CRG--921322 (LD and DAK) and CRG--910285
(ZB and DCD).

\vfill
\break
\listrefs

\bye